\def\lans{l }
\ifx\answ\lans
\documentstyle{article}

\catcode`@=11
\def\jimtwoup{\twocolumn\sloppy\flushbottom\parindent 2em
        \parskip .33\baselineskip
        \leftmargini 2em\leftmarginv .5em\leftmarginvi .5em
        \oddsidemargin 0in      \evensidemargin 0in
        \columnsep .4in \footheight 0pt
        \textwidth 10in \topmargin  -.4in
        \headheight 0pt \topskip 0in
        \textheight 6.9in \footskip 30pt
        \hoffset -.5in \voffset -.25in
        \def\@oddfoot{\hfil\thepage\hfil\addtocounter{page}{1}
                \hspace{\columnsep}\hfil\thepage\hfil}
        \let\@evenfoot\@oddfoot \def\@oddhead{} \def\@evenhead{} }
\jimtwoup
\else
\documentstyle[12pt]{article}
\def\hybrid{\topmargin 0pt      \oddsidemargin 0pt
        \headheight 0pt \headsep 0pt
        \textwidth 6.5in        % US paper
        \textheight 9in         % US paper
        \marginparwidth .875in
        \parskip 5pt plus 1pt   \jot = 1.5ex}
\catcode`@=11
\hybrid
\fi
\catcode`@=12

\catcode`\@=11
\def\marginnote#1{}
\newcount\hour
\newcount\minute
\newtoks\amorpm
\hour=\time\divide\hour by60
\minute=\time{\multiply\hour by60 \global\advance\minute by-\hour}
\edef\standardtime{{\ifnum\hour<12 \global\amorpm={am}%
        \else\global\amorpm={pm}\advance\hour by-12 \fi
        \ifnum\hour=0 \hour=12 \fi
        \number\hour:\ifnum\minute<10 0\fi\number\minute\the\amorpm}}
\edef\militarytime{\number\hour:\ifnum\minute<10 0\fi\number\minute}
\def\draftlabel#1{{\@bsphack\if@filesw {\let\thepage\relax
   \xdef\@gtempa{\write\@auxout{\string
      \newlabel{#1}{{\@currentlabel}{\thepage}}}}}\@gtempa
   \if@nobreak \ifvmode\nobreak\fi\fi\fi\@esphack}
        \gdef\@eqnlabel{#1}}
\def\@eqnlabel{}
\def\@vacuum{}
\def\draftmarginnote#1{\marginpar{\raggedright\scriptsize\tt#1}}
\def\draft{\oddsidemargin -.1truein
        \def\@oddfoot{\sl preliminary draft \hfil
        \rm\thepage\hfil\sl\today\quad\militarytime}
        \let\@evenfoot\@oddfoot \overfullrule 3pt
        \let\label=\draftlabel
        \let\marginnote=\draftmarginnote
   \def\@eqnnum{(\theequation)\rlap{\kern\marginparsep\tt\@eqnlabel}%
\global\let\@eqnlabel\@vacuum}  }
\def\numberbysection{\@addtoreset{equation}{section}
        \def\theequation{\thesection.\arabic{equation}}}
\def\underline#1{\relax\ifmmode\@@underline#1\else
        $\@@underline{\hbox{#1}}$\relax\fi}
\def\titlepage{\@restonecolfalse\if@twocolumn\@restonecoltrue\onecolumn
     \else \newpage \fi \thispagestyle{empty}\c@page\z@
        \def\thefootnote{\fnsymbol{footnote}} }
\def\endtitlepage{\if@restonecol\twocolumn \else  \fi
        \def\thefootnote{\arabic{footnote}}
        \setcounter{footnote}{0}}  %\c@footnote\z@ }
\catcode`@=12
\relax
\def\ie{\hbox{\it i.e.}}

\def\beq{\begin{equation}}
\def\eeq{\end{equation}}
\def\bea{\begin{eqnarray}}
\def\eea{\end{eqnarray}}

\let\vev\VEV
\relax
\numberbysection
\begin{document}
\begin{titlepage}
\begin{center}
February~1995 \hfill    PAR--LPTHE 9513, SU-HEP 4241-608, EFI 95-17 \\
        \hfill hep-th/9xmmmnn\\[.5in]
{\large\bf Self-Avoiding Surfaces in the 3{--}$d$ Ising
Model }\\[.5in]
        {\bf Vladimir~S.~Dotsenko\footnote{Also at the Landau Institute for
Theoretical Physics, Moscow}, Marco~ Picco and  Paul~Windey}\\
        {\it LPTHE\/}\footnote{Laboratoire associ\'e No. 280 au CNRS}\\
        {\it  Universit\'e Pierre et Marie Curie - PARIS VI\\
        Universit\'e Denis Diderot - Paris VII\\
        Boite 126, Tour 16, 1$^{\it er}$ \'etage \\
        4 place Jussieu\\
        F-75252 Paris CEDEX 05, FRANCE}\\[0.5em]
        {\bf Geoffrey~Harris\footnote{present address:  Enrico Fermi
         Institute and Department of Physics, University of Chicago}}\\
        {\it Physics Department, Syracuse University\\
        \it Syracuse, NY 13244, USA}\\[0.5em]
        {\bf Emil~Martinec}\\
        {\it Enrico Fermi Institute and Department of Physics,\\
        University of Chicago\\
        Chicago, IL 60637, USA}\\[0.5em]
        {\bf Enzo~Marinari}\\
	{\it Dipartimento di Fisica and INFN\\
        Universit\`a di Cagliari,\\
	Via Ospedale 72, 09100 Cagliari, ITALY}\\[0.5em]
\end{center}
\vskip .5in
\newpage
\centerline{\bf ABSTRACT}
\begin{quotation}
We examine the geometrical and topological properties of surfaces
surrounding clusters in the 3--$d$ Ising model.  For geometrical clusters
at the percolation temperature and Fortuin--Kasteleyn clusters at $T_c$,
the number of surfaces of genus $g$ and area $A$ behaves as
$A^{x(g)}e^{-\mu(g)A}$, with $x$ approximately linear in $g$ and $\mu$
constant.  These scaling laws are the same as those we obtain
for simulations of 3--$d$ bond percolation.  We observe that
cross--sections of spin domain boundaries at $T_c$ decompose into a
distribution $N(l)$ of loops of length $l$ that scales as $l^{-\tau}$ with
$\tau \sim 2.2$. We also present some new numerical results for 2--$d$
self-avoiding loops that we compare with analytic predictions. We address the
prospects for a string--theoretic description of cluster boundaries.
\end{quotation}
\end{titlepage}
\newpage
\section{Introduction}

     One of the major successes of 20th century physics has been the
expression of the critical behavior of a variety of theories of nature
in terms of sums over
decorated, fluctuating paths.  It has thus been hoped that
higher dimensional analogues, theories of fluctuating membranes, also
play a fundamental role in characterizing the physics of critical phenomena.
In particular, significant effort has been invested in
recasting one of the simpler models of phase transitions, the
3{--}$d$ Ising model \index{Ising model!in 3 dimensions},
as a theory of strings \cite{string} \index{String theory}.  These
attempts have
been stymied by the
difficulty in taking the continuum limit of formal sums over lattice
surfaces.

     In fact, sums over lattice surfaces, built from e.g.
plaquettes or polygons, generically fail to lead to a well-defined continuum
theory of surfaces.  An exception to this rule occurs when
the surface discretizations are embedded in $d \leq 1$.
In this case, one can exactly solve a large class of toy lattice models
which lead to sensible continuum `bosonic' string theories
(at least perturbatively) \cite{clt1}.  Numerically, it is observed that the
$d > 1$ versions of these
lattice models suffer a `fingering instability'; the embedded surfaces,
for instance are composed of spikes with thickness of the order of the
cutoff. It is suspected that the polygonal discretization of the worldsheet
(for large volumes)
is configured in a polymer{--}like structure, so that these theories
cannot be realized as sums over surfaces in the
continuum limit\index{Branched Polymers}.
This instability is anticipated theoretically, since the mass--squared
of the dressed identity operator of the bosonic string
becomes negative above $d=1$, presumably
generating a uncontrolled cascade of states that tear
the worldsheet apart\cite{tachyon}\index{Tachyon}.

    In the continuum limit, we know how to evade these problems
in special cases
through the implementation of supersymmetry and the GSO projection.  This
additional structure, however, leads to fundamental difficulties in
discretizing these theories.  In principle, one might hope to somehow
guess an appropriate continuum string theory and then show that it embodies
the critical behavior of a lattice theory, such as the 3--$d$ Ising
model.  The prospects for success through such an approach
seem rather poor at this time.

      Given this state of affairs, we have turned to a more
phenomenological approach, in which we attempt to generate `physical'
random surfaces in a particular model and then examine their
topological and geometrical properties.  We thus have chosen to look
at the structure of domain boundaries in the 3{--}$d$ Ising model.
The phenomenology of these self{--}avoiding
cluster boundaries is interesting in its own
right, since it describes a large universality class of behavior that is
expressed frequently and quite precisely by nature.  We
also might hope that our observations may be useful in gauging the
prospects of success of a string--theoretic description.  The
Ising model has been employed previously as a means to generate random
lattice surfaces \footnote{Through the use the phrase `lattice
surface' rather than `surface', we indicate that these objects should
{\it not} be necessarily inferred to be
real surfaces in the continuum limit.};
see for instance, the work of David \cite{david2}, Huse and Leibler
\cite{HuseLei}, Karowski and Thun
and Schrader \cite{previous}.  In a sense, this work extends these studies
by looking for new features of the geometry of these lattice surfaces; we also
consider boundaries of Fortuin--Kasteleyn clusters as well as
`geometrical' spin domains.  Much of our analysis consists of a measurement
of the distribution of surfaces as a function of their area $A$ and genus
$g$, $N_g(A)$
\footnote{The mean genus per Ising configuration is measured in references
\cite{previous}.  A determination of genus as a function of
area in an Ising system with anti--periodic boundary conditions has
been made \cite{Gliozzi}.}.   We shall determine the
functional form of $N_g(A)$.  We also perform block spin measurements
of the genus, to determine if a condensation of handles is present
on cluster boundaries at all scales.  These cluster boundaries
are strongly coupled and thus it appears cannot be directly characterized
by perturbative string theory.   We see that, however, boundaries of spin
domains at the Curie temperature are not just strongly--coupled
versions of the branched polymer--like objects that attempts to
build `bosonic' random surfaces typically generate.
They instead exhibit a richer
fractal structure,
albeit one not characteristic of surfaces.
We show that they obey a new scaling law
that describes the
distribution $N(l)$ of lengths $l$ of loops that compose cross--sections of
cluster boundaries.

     In the course of these investigations, we generated a considerable
amount of data characterizing additional geometric properties of
Ising clusters and their bounding surfaces.  In particular, we also
simulated the two--dimensional Ising model.  In the two--dimensional
case, there exist many exact results describing
the fractal structure and distribution of clusters and loops.   To provide
a more comprehensive and complete picture of the geometry of Ising surfaces,
we shall present these additional results in this paper.  In some cases,
these additional results have been verified previously, though generally
on smaller lattices and with somewhat less numerical precision.

\section{Ising Clusters and Surfaces}

      We shall begin by summarizing the basic physical properties
of the cluster boundaries that we have analyzed.
To a first approximation, a 2{--}dimensional membrane
of area $A$ and curvature matrix $K$
will exact an energy cost \cite{HuseLei,David}
\begin{equation}
\label{action}
H = \mu A + \lambda \int ({\rm{Tr}}K)^2 + \kappa \int {\rm{Det}}K;
\end{equation}
$\mu$ is the bare surface tension, $\lambda$ is referred to as the
bending rigidity and $\kappa$ couples to the Euler character of the
surface.  In the regime which characterizes random surfaces\index{Random
Surfaces},
the surface tension must be sufficiently small to allow
significant thermal fluctuations.
Note that the above action does not constitute a complete physical
description of the Ising surfaces.
It is essential also to keep in mind the constraint that
Ising cluster boundaries are naturally
self--avoiding.
We first consider
surfaces in the dual lattice that bound `geometrical clusters' formed
from sets of adjacent identical spins.  In this case, the Ising dynamics
generates an energy penalty proportional to the boundary area;
$\lambda_{bare}$ and $\kappa_{bare} = 0$.  The bare surface tension is
tuned by the Ising temperature.   To put this model in perspective,
we note that for real vesicles, for instance,
the couplings
$\lambda$ and $\kappa$ can be quite large; $\lambda$ ranging from
about $kT$ to $100kT$ have been measured \cite{David}.  The bending
rigidity may be irrelevant in the continuum limit, however.
The string coupling \footnote{We ignore
distinctions between intrinsic and extrinsic metrics.} is
equal to $\exp (-\kappa)$.
Through blocking spins, we make an estimate of the renormalization
group behavior of $\kappa$.
Unless $\kappa$ effectively becomes large in the infrared, the cluster
boundaries will fail to admit a surface description in the continuum limit.

       The geometrical clusters and their boundaries are {\it{not}}
present at all scales at the Curie
temperature\index{Clusters!Ising}.  Instead, for
temperatures somewhat below $T_c$ and all temperatures above $T_c$
two huge geometrical clusters comprise a finite fraction
of the entire lattice
volume.   These clusters percolate, that is, they wrap around
the entire lattice (we shall consider periodic boundary conditions).
Otherwise, the lattice only contains very small clusters that are the
size of a few lattice spacings; there are no intermediate size clusters.
We can understand this behavior by considering the $T \rightarrow
\infty$ behavior of these clusters.  Two percolated clusters span
the lattice even at infinite temperature, where clusters are smaller
than at $T_c$.  At $T = \infty$, the spins are distributed randomly
with spin up with probability $50\%$; the problem of constructing
clusters from these spins then reduces to pure site percolation with
$p=1/2$.  Pure site (or bond) percolation describes the properties of
clusters built by identifying adjacent colored bonds (sites), which are
colored randomly with probability $p$\index{Percolation}.  Above a critical
value $p=p_c$, the largest of these clusters percolates through
the lattice\cite{Stauffer}.
For the cubic lattice, it is known that an infinite cluster
will be generated (in the thermodynamic limit)
at $p_c \sim .311$. Thus, the fact that
the geometrical clusters have percolated in the high--temperature
regime and at the Curie point is
essentially a consequence of the connectivity of 3--$d$ lattices.

    At very low temperatures, however, there are few reversed (minority) spins
in the Ising model; these form a few small clusters.  As the density of
minority spins increases, the clusters become bigger until the largest
cluster percolates at some temperature $T_p < T_c$.
It has been suggested (see \cite{CambNaue} and
\cite{HuseLei})
that since this minority spin percolation appears to be due to an increase
in the concentration of minority spins and not to any long--distance
Ising dynamics, that this transition is in the same universality class
as pure (bond or site) percolation.  We emphasize that the scaling
of minority clusters
should not correspond to any non--analyticity in the thermodynamic
behavior of the Ising model; it should essentially be a `geometric effect'.

      There is another type of cluster, introduced by Fortuin and
Kasteleyn \cite{FK,CK}, that does proliferate over all length scales at the
Curie point\index{Clusters!Fortuin--Kasteleyn}.  These FK clusters
consist of sets of bonded spins; one draws these bonds between
adjacent
same--sign spins with a temperature dependent probability $p = 1 -
\exp(-2\beta)$.  Note that the geometrical clusters are built by
a similar procedure, using instead $p=1$.  FK clusters
arise naturally in the reformulation
of the Ising model as a percolating bond/spin model \cite{Sokal}.
For the Ising partition function can be rewritten as a sum over occupied and
unoccupied bonds with partition function
\begin{equation}
 Z = \sum_{bonds}p^{b}(1-p)^{(N_b - b)}2^{N_c}
\protect\label{BSPART}
\end{equation}
where $p = 1-\exp(-2\beta)$, $N_b$ denotes the number of bonds in
the entire lattice in which $b$ bonds are occupied and $N_c$ equals
the number of clusters that these occupied bonds form.  When the factor
$2^{N_c}$ is replaced by $q^{N_c}$, then (\ref{BSPART}) is the partition
function for the $q$-state Potts model.  If we
assign a spin to each bond so that all bonds in the same cluster
have the same spin, then the factor of $q^{N_c}$ just comes
from a sum over spin states.  The above partition function can then be viewed
as a sum over FK clusters.  Using this
construction, one can show that the spin-spin correlator in the original
Ising model is equal to the  pair connectedness function of FK clusters,
\begin{equation}
\langle \sigma(x)\sigma(y) \rangle = \langle \delta_{C_x,C_y}
\rangle ,
\protect\label{SSREL}
\end{equation}
which equals the probability that points $x$ and $y$ belong to
the same FK cluster \cite{Hu}.  It then follows that for $T \ge T_c$,
the mean volume of the FK clusters is proportional to the susceptibility
of the Ising model, so that indeed FK clusters only just start to
percolate at the Curie point.  Additionally, the relation (\ref{SSREL})
also implies that the spatial extent of the FK clusters is proportional to
the correlation length of the Ising model.
Furthermore, scaling arguments \cite{Wang}
demonstrate that at $T_c$, the volume distribution of FK clusters obeys
\begin{equation}
\label{vscaling}
N(V) \simeq V^{-\tau}, ~~~  \tau = 2 + \frac{1}{\delta},
\end{equation}
where $\delta$ denotes the magnetic exponent of the Ising model
($M \simeq B^{1/\delta}$).  Thus we see that FK clusters, unlike
the geometrical clusters previously discussed, directly
encode the critical properties of the Ising model.  Indeed, we are
necessarily led to study FK clusters in order to measure scaling laws
that characterize cluster boundaries of the scale of the Ising
correlation length, i.e. boundaries that scale at the Curie point.
On the other hand, geometrical cluster boundaries contribute an
energy penalty proportional to their individual area; the lattice
surface dynamics of FK cluster boundaries, however, cannot be likewise
described by a similar physical rule.

In 2--dimensions both the FK clusters and
the geometrical clusters
percolate at the Curie temperature.  The critical properties
of these clusters differ, however, since the scaling of geometrical clusters
is partially determined by the `percolative' properties of
two--dimensional lattices.  These effects are in some sense removed through
the FK construction. We will present below numerical results for
2--$d$ geometrical clusters, which can be compared with theoretical
predictions\cite{a1,cardy}.

\section{The Simulation}

We now proceed to outline the techniques used in our Monte Carlo
simulations\index{Monte--Carlo simulations}. We performed a set of
medium-sized simulations using about one year of time on RISC workstations.
We collected data on a variety of two and three-dimensional lattices:
square, triangular, simple cubic and BCC (see below). A third
set of measurements of distributions of loop sizes was made on
two-dimensional slices of three-dimensional lattices. A summary
of the size of our runs appears in tables 1-3.

Spin updates were implemented through the efficient Swendsen--Wang
algorithm \cite{SW}:  FK clusters for each lattice configuration
are first constructed, then the spins composing each cluster
are (all) assigned a new random spin value.

The main technical difficulty that we encountered (in three dimensions) was
the measurement of the Euler character, equal to $V - E + F$ for a dual
surface with $V$ vertices, $E$ edges and $F$ faces.  On the simple cubic
(SC) lattice, the construction of the dual surface and measurement
of genus is ambiguous.   Each surface is built from plaquettes composing
the phase boundary between a pair of clusters, e.g. cluster {\it a}
and cluster {\it b}.  One can then associate with this surface the
set of cubes in the dual lattice that surround sites in cluster {\it a}
along the surface boundary.  To measure genus we must then
resolve two types of ambiguities in building these surfaces.  These
ambiguities occur when the associated cubes intersect along just
one link or intersect only at a vertex.  One has to decide, for example,
whether to connect cubes that touch at just a vertex with a thin tube
or to instead, split them, so they no longer touch.  We came up with three
separate algorithms (two of which turned out to be equivalent)
that are consistent in the following sense:
they yield the standard value of genus when no ambiguities were present
and they always lead to a genus that is a non--negative integer.  We
chose, for instance, to split cubes that touched at just one point.
A consistent algorithm to measure genus on the simple cubic lattice
is also presented in the work of Caselle, Gliozzi and Vinti\cite{Gliozzi}.

  Since these rules are not unique, one
would hope that their implementation essentially serves as a regularization
that does not affect long--distance scaling laws. In fact, in order to
eliminate any doubts about our rules, we also performed simulations on
specially chosen lattices where ambiguities are absent.
In two dimensions, one can avoid ambiguous
intersections on the dual
lattice by considering Ising spins on the triangular lattice.  Its dual
(the honeycomb lattice) is trivalent and thus Ising spin domains will not
be enclosed by self--intersecting paths.  This fortuitous situation
generalizes to three--dimensions for the Ising model on a body centered
cubic (BCC) lattice in which the vertices at the center of each cube are
also connected to those in the centers of neighboring cubes.  More
explicitly, we coupled with equal strength both the $6$ nearest and $8$
next-nearest Ising spins so that only three plaquettes of the dual lattice
meet along a dual link.  Since surfaces built dual to this lattice are also
naturally self-avoiding, computing the genus is trivial.  A depiction of
the Wigner--Seitz cell of this lattice (composed of plaquettes in the dual
lattice) appears in figure 3.1.

   In two dimensions, measurements were performed at the Curie
temperature.  Configurations of FK clusters and cross-sectional
slices were taken at the three-dimensional Curie temperature.
For the SC lattice, this value is well known \cite{Hasenbuch}.
On the BCC lattice with second nearest neighbor interactions, we determined
the Curie temperature by adjusting $\beta$ until we found optimal
scaling for the cluster size distribution.  In three dimensions,
we examined the scaling of geometrical clusters at the percolation
temperature $\beta_P$.  We determined this using a method discussed
by Kirkpatrick \cite{kirk} in which one measures the fraction
of configurations $f$ containing clusters that span the lattice as
a function of $\beta$.  One plots $f$ versus $\beta$ for different
lattice sizes $L$; $\beta_P$ corresponds to the intersection of these
curves for different $L$.

Statistical errors are computed using binning and the jacknife
technique.  We determine exponents through linear least-squared fits;
statistical errors for these exponents are also obtained by using jacknife
when fitting.  Generally, systematic errors swamp our statistical
errors.  These systematic effects are due to finite-size effects,
the failure to reach the asymptotic scaling region as well as the
uncertainty in the
value of the critical temperature in certain cases.  The absence of a
quoted error or an errorbar henceforth indicates that the statistical error
is much smaller than our measured observable or that the errorbars are too
small to appear on our plots.  In particular cases (when we examine slices
of $150^3$ lattices), our data will not be sufficient to accurately
estimate the jacknife error.  We are confident in these cases, though,
that the statistical error is still much smaller than the systematic error.

\section {Clusters and self-avoiding loops in the 2D Ising model}
\label{2dresults}
In order to check our methods and techniques we first turn to the $2$-$d$
Ising model. In fact, in two dimensions, a large number of
critical exponents have
been computed by using conformal field theory techniques \cite{a1,cardy}.
We shall see that our measurements agree with these predictions.
We determined the scaling properties of geometrical clusters
and of self-avoiding loops bounding these clusters on square and
triangular lattices with sizes up to $1000 \times 1000$.  The measured
scaling laws and lattices were chosen for
their similarity to the three dimensional analogues
that we are most interested in.  In particular,
the honeycomb lattice (dual to the triangular lattice) is well known to
produce self-avoiding loops in a natural way since it has a coordination
number equal to three.  It is analogous in this respect to the dual of the
BCC lattice in three dimensions. For both the triangular lattice
and the BCC lattice,
there exists no ambiguities in defining the boundary of a spin cluster.

The equilibrium configurations were produced by a Swendsen--Wang cluster
algorithm at $T_{c} = 0.44068\ldots$ for the square lattice, and at $T_{c}
= 0.27465\ldots$ for the triangular lattice. This algorithm is supposed to
have a relaxation time exponent equal to zero precisely for the 2-$d$ Ising
model \cite{sw}.  After every $10$ cluster updating steps we analyzed the
resulting spin configuration.  For each of these configurations we measured
$N(l)$, the statistical distribution of the self-avoiding loops bounding
the spin clusters, as a function of the length of their perimeter.  We also
measured $A(l)$, the average total area inside these loops.

The definition of what we call $A(l)$ needs to be made precise: by area
$A(l)$ we mean the {\em total} area enclosed by a given loop of length $l$.
This area includes the spin cluster bounded by the loop of length $l$;
it also incorporates all of the islands of flipped spins imbedded within
this cluster.
We consider all loops, not just the
outer ones and each given loop is considered as an outer boundary or
``hull'' of the complete figure inside.
We used this definition since it
appears to be the most natural for the problem of self-avoiding loops.
It makes sense to consider all loops, since on an infinite lattice, any
given loop would be inside other larger loops at $T_c$.

We define the   exponents $\tau$ and $\delta$ by
\beq
N(l)\simeq l^{-\tau}
\label{L1}
\eeq
and
\beq
A(l)\simeq l^{\delta}.
\label{L2}
\eeq
The values that we obtained for $\tau$ and $\delta$ are listed
in tables 4 and 5.
The windows (intervals of $l$) were chosen as usual to minimize the influence
of corrections to scaling at small $l$ and
finite-size effects at large $l$.
The errors that we quote for these exponents reflect the
systematic uncertainty arising from our choice of windows.
These systematic errors should be larger than the statistical uncertainties,
which nonetheless are difficult to estimate.

Our best results were obtained with the $1000 \times 1000$ triangular
lattice. They give the following scaling exponents for self-avoiding loops
in two dimensions: \beq \tau = 2.44 \pm 0.01, \quad \quad \quad \delta =
1.454 \pm 0.002.
\label{L3}
\eeq
The remarkable scaling behavior of $N(l)$ and $A(l)$ is displayed (in
$\log-\log$ plots) respectively in figure 4.1 and figure 4.2.

The values of the exponents in (\ref{L3}) can be compared with the
theoretical predictions based on the Coulomb gas representation \cite{a1}
and with further scaling arguments originally due to B.~Duplantier
\cite{a2} \footnote{For an alternate derivation of a similar relation in
the case of percolation theory, see section 3.4 of Stauffer's
book\cite{Stauffer}}.\marginnote{should  we give here a discussion of these
arguments? \\PW.} This theoretical analysis yields the scaling relation:
\beq
\tau = 1 + \delta
\label{L4}
\eeq
with a value for $\delta$ of
\beq
\delta = \frac{2}{D_{H}} = \frac{16}{11} = 1.4545 \ldots .
\label{L5}
\eeq
Here $D_{H} = \frac{11}{8}$ \cite{a1}
is the fractal dimension commonly used for
the cluster ``hulls''.\marginnote{give the def. of $D_H$?\\PW}
One observes that our numerical values  (\ref{L3}) are in good agreement
with these theoretical predictions.   We shall present a version of
Duplantier's derivation of
the relation between $\tau$ and $\delta$ in the appendix.

Previous numerical work on other exponents related to $D_{H}$ can
be found in \cite{a3,a4}.  The results of these papers support the
theoretical value of $D_{H}$ given above.  One should remark that our
simulations, which are done on a much larger lattice ($1000^2$ rather than
$36^2$ as  in \cite{a4}), yield much more accurate values of the exponents.

Finally, we measured on the triangular lattice the universal ratio
\beq
{A(l) \over R^2(l) } = r_l.
\eeq
$r_l$, which was recently computed by Cardy \cite{cardy}, is the ratio of
the area inside a loop of length $l$ to the squared radius of gyration of
this same loop, defined by
\beq
R^2 ={1\over 2 l^2}\sum_{r_1,r_2} (r_1-r_2)^2
\eeq
($r_1, r_2$ are the positions of the links of the loop on the lattice).
The result obtained by Cardy is
\beq
r_l \sim {{1+2g}\over{2(1+2g)}}\pi
\eeq
where $g$ is a Coulomb gas parameter with $g={4\over 3}$ for Ising
clusters at $T_c$ and $g={2\over
3}$ for
Ising clusters at $T=\infty$ (which in fact corresponds to the pure
percolation point on the triangular lattice; this point is discussed in
section 5.3). The values that we obtained for $r^c_l$ ($r_l$ at the
critical temperature) and $r^\infty_l$ ($r_l$ at $T=\infty$) are listed in
table 6. Here again, intervals were chosen so as to avoid lattice
artifacts at small $l$ and finite-size effects at large $l$.
 From our measurements, we
deduce, for the $1000 \times 1000$ lattice, the following value at the
critical temperature:
\beq
r^c_l = 2.471\pm 0.001;
\eeq
the exact value given by Cardy is $r^c_l={11\over 14}\pi \simeq
2.468\ldots $.
The measured quantity again approaches the exact value as we increase
the lattice size. On $500\times 500$ lattices, the measured value is
$r^c_l=2.472\pm 0.001$ while on $250 \times 250 $ lattices, which we also
simulated, it is $r^c_l=2.478\pm 0.002$.

For the percolation case (\ie $\;T=\infty$), the value given by Cardy is
$r^\infty_l={7\over 10}\pi \simeq 2.199\ldots $. We
obtain, for the $1000 \times 1000$ lattice, the value
\beq
r^\infty_l=2.218\pm 0.001.
\eeq
This differs from Cardy's prediction by about
$1$ percent. But again, as we increase the lattice size, the
measured value approaches the exact value.  For the
percolation case, lattice artifacts at small $l$ are important up to a
value of $l$ of order $\simeq 200$.
We are able to obtain only limited statistics in the
regime that exhibits good asymptotic behavior.
\section{Results for 3D Clusters}
We now present data from our simulations on both the simple cubic and BCC
lattices.  We have examined boundaries of FK clusters at $T_c$, surfaces
bounding minority spin domains at $T_p$, geometrical clusters at $T_c$ and
pure bond percolation. A more concise summary of some of these results has
been presented in \cite{ourletter,proceedings}.

\subsection{Cluster Geometry}

We begin by discussing geometrical properties of the clusters.  Some of the
material in this section is already well known, but we present it to
illustrate the influence of lattice artifacts and finite-size effects in
our data.  This analysis will allow us to determine the range of parameters
for which we will be able to best trust our results.

First, we shall analyze the data for FK clusters on the simple cubic
lattice with volumes $32^3$ and $64^3$.  We fit to the cluster distribution
function

\begin{equation}
  N(V) \simeq  V^{-\tau}\ ,
  \protect\label{NV}
\end{equation}
where $V$ is the cluster volume in real space (see figure 5.1).

As a first check we have reproduced the exact fit used by Wang in
ref. \cite{Wang}, by using
volumes ranging from $4$ to $64$.  Wang quotes here a value of $2.30$
(against the expected value of $2.21$ \footnote {In particular, by applying
the scaling relations to the results of $\epsilon$ expansions, one expects
$\tau = 2.207 (1)$ \cite{Bake}, high temperature expansions yield
$\tau = 2.210 (1)$ \cite{DombZinn2} and RG
calculations give $\tau = 2.207(<1)$ \cite{ItzyDrou}.
}).

Our best fits for $V$ ranging from $4$ to $64$ (which we present here
only for the sake of comparison) give $\tau^{FK}_{L=32}=2.310$,
$\tau^{FK}_{L=64}=2.299$, where in both cases the statistical error is less
than one in the last digit.  Note that $\tau^{FK}$ decreases slightly as a
function of increasing lattice size.  It is evident that the lattice
sizes we used are not sufficient to exclude both significant corrections
to scaling (for small $V$) or finite-size effects (afflicting
$V \sim L^3$).   We do not observe a convincing plateau in plots of
$\log N(V)$ vs. $\log V$.
The closest the data comes to plateauing on $L=32$ lattices
is in the volume range $V \in (32,1024)$, where we extract
$\tau^{FK}_{L=32} = 2.324 \pm .001$.  We obtain $\tau^{FK}_{L=64} =
2.286 \pm .001$ on $L=64$ lattices.

Likewise, on the $BCC$ lattice, we see large deviations from
power--law scaling of $N(V)$.  In this case,
our values of $\tau_{FK}$ are quite close to the theoretical prediction
of $2.21$; for $L=64$ we measure $\tau^{FK} = 2.235$ and $2.218$ on
the volume windows $(32,1024)$ and $(64,1024)$ respectively.  This agreement
with theory should be viewed with a great deal of caution, given
the large systematic effects that are present.

     Similar results hold in the analysis of $N(V)$ for geometrical clusters
at $T_p$.  Significant deviations from scaling are again present.  We
measure smaller values of $\tau$ than in the FK case:  $\tau^{GC}_{L=30}
= 2.069 \pm .005$, $\tau^{GC}_{L=60} = 2.124 \pm .002$ and
$\tau^{GC}_{L=100} = 2.13 \pm .002$ on windows of size $(32,256)$,
$(64,2048)$ and $(64,4096)$ respectively.
     One would anticipate that the value of $\tau$ for geometrical clusters
would be characteristic of the scaling of pure percolation clusters.
For pure bond percolation, the scaling exponents have been determined
primarily through series expansions and to a lesser extent through
Monte Carlo techniques; these analyses give a value of $\tau$
that is centered about $2.18$ with an uncertainty of roughly $0.02$
\cite{Stauffer}.

   We also measure $N(V)$ explicitly for $3d$ bond percolation; in this
case, the data are much cleaner.
On the $32$ to $1024$
window, for example, we get an exponent of $2.217\pm 0.001$ both on the
$32^3$ and $64^3$ lattice. On the $128$-$2048$ window we see the first
(small!) signs of finite size effects, with a value of $2.200\pm 0.004$
for the small lattice and $2.209\pm 0.002$ on the large lattice. The
results are so precise and consistent in this case that we can attempt
a fit
to finite size corrections; this yields a
result in between $2.18$ and $2.21$, in complete agreement with
the numbers cited in the literature \cite{Stauffer}.
We stress that this extrapolated
number apparently has a much smaller systematic error than the ones we
have quoted in other cases.

It is thus evident that the power law fits to $N(V)$ are a
rather poor way to measure critical exponents; much more accurate estimates
can be obtained through finite--size scaling fits of the mean cluster size as a
function of lattice size $L$\index{Finite--size scaling}.  We now present
our finite-size scaling analysis.
First, we have measured the scaling behavior
(from $L=32$ to $L=64$) for the mean cluster size
\begin{equation}
  \frac{\langle V^2 \rangle}{\langle V \rangle} \simeq L^{\cal{H}}\ ,
\end{equation}
finding an exponent of $1.97\pm 0.01$ for FK Ising clusters on the SC
lattice, an exponent of $1.99\pm 0.01$ for geometrical clusters on the BCC
lattice (where we extrapolate from $L=30$ to $L=100$), and an
exponent of $2.09 \pm 0.01$ for bond percolation
\footnote{  Note that
we ran our bond percolation simulations at $p = .249$; recent Monte
Carlo work indicates that actually $p_c$ may be as low as $.2488$
\cite{Adler} in this case.  We would then estimate (by noting how
sensitive $\cal{H}$ is to $p$) that the uncertainty in $p_c$ contributes
to a systematic error of roughly $0.02 - 0.03$ in $\cal{H}$ for percolation.
Since the Curie temperature is known much more precisely, this
bias is not significant for FK clusters.}.
Since the mean cluster size is proportional to the susceptibility,
it obeys the finite-size scaling relation characteristic of the
susceptibility at $\beta_c$, so $\cal{H} = \frac{\gamma}{\nu}$.
For the $3d$ Ising model, predictions for $\frac{\gamma}{\nu}$ are
$1.97(1), 1.95(1)$ and $1.97(1)$ from $\epsilon$ expansions, high temperature
series and renormalization group calculations respectively.
Series expansions, Monte Carlo simulations and $\epsilon$
expansions have been also applied to the calculation of pure percolation
exponents.  In this case, they have yielded the values
$\frac{\gamma}{\nu} = 2.07(16), 2.05(2)$ and
$2.19(11)$ respectively.  Our measurements of the finite-size
scaling behavior of the mean cluster size thus appear to yield precise
and correct estimates of $\frac{\gamma}{\nu}$. Now, using some standard
scaling relations and (\ref{vscaling}) it is also
possible to relate ${\gamma\over \nu}$ to $\tau$ and then obtain a second
measurement of $\tau$ :
\beq
\tau = (3 + \gamma/\nu d)/(1 +
\gamma/\nu d)~~~~ (d = 3)
\eeq
Using this technique, we
measured $\tau_{{\rm FK}} = 2.207(3)$
on the SC lattice
and $\tau_{{\rm geo}} = 2.202(3)$ on the BCC lattice.  The error
on $\tau_{{\rm geo}}$ is in fact probably several times larger than
quoted above, due to
uncertainties in locating the critical temperature.
This measurement of $\tau_{{\rm FK}}$ agrees
perfectly with previous values; the measurement of $\tau_{{\rm geo}}$
is not accurate enough to distinguish likely pure percolation
behavior from that of percolation of FK clusters.

We have also measured the exponent given by the scaling of the maximal
cluster volume, defined by
\begin{equation}
  \langle V_{Max} \rangle \simeq L^{\cal{J}}\ ,
\end{equation}
finding $2.49\pm 0.01$ for SC FK Ising clusters, $2.53\pm 0.01$ for
BCC geometrical clusters and $2.56\pm 0.01$ for bond
percolation.   One can show via scaling arguments (from the relation
(\ref{NV})) that ${\cal{J}} = \frac{3}{\tau - 1}$ and then applying
standard scaling relations, that ${\cal{H}} = 2{\cal{J}} - 3$.  We thus
see that our values of $\cal{J}$ are consistent with those of $\cal{H}$.

To get a better picture of the cluster geometry, we also examined
the dependence of the cluster surface extent $A_c$ on
its volume $V$.  Note that for $V < 6$, the simple cubic lattice
structure demands $V$ = $A_c$.  For slightly higher volumes,
clusters do begin to form interior points, so that $A_c$ becomes less
than $V$.

It is well known that typical pure percolation clusters
are saturated with holes and
crevasses which break up their scant interiors.  Since unoccupied bonds
are distributed homogenously
with probability $1-p$, there is a fixed probability per unit area
that any site will not be pierced by occupied bonds, but that
its neighboring site will belong to a cluster.
 From this argument one can deduce
\cite{Leath} that the cluster perimeter
(defined as the number
of empty sites adjacent to an occupied cluster site) is linearly proportional
to the cluster volume; percolation clusters are tubular and very
branched.

 Note that FK clusters are
formed by implementing pure bond percolation on geometrical
Ising clusters.  Therefore, one might anticipate that they at least
qualitatively might share some of the geometrical characteristics of pure
percolation clusters.  In particular, one could argue that their
perimeter should be linearly proportional to their volume by applying
the above reasoning.
Indeed, in all cases (FK, geometrical and pure percolation clusters),
we found that the cluster perimeter was proportional to the enclosed volume.
This dependence can be characterized by the effective exponent $\omega$,
given by
\begin{equation}
        A_c \sim V^{\omega}\ .
\end{equation}
For instance, we find that on the $64^3$ lattice, in the window $49 < V < 293$,
$\omega^{FK} = 0.980$; $\omega$ steadily grows as $V$ increases until
it reaches $0.992(1)$ in the window $611 < V < 841$.

\subsection{Cluster Topology}

   We have computed the quantity $N_g(A)$ (the number of dual surfaces of
given genus $g$ and area $A$) for the models that we have
studied. Our data clearly show
that we can model scaling laws for such complex
quantities.
It turns out that in all cases our data
are described asymptotically by
\beq
  N_g(A) = C_gA^{x(g)} e^{-\mu(g) A}\ ,
  \protect\label{ENG}
\eeq
we will discuss the cases in which, due to lattice
artifacts and finite size effects, this behavior is not perfect.
We have used in this formula a generic genus $g$ dependence
$x(g)$ and $\mu(g)$, but we will argue that our data suggests
that asymptotically $\mu$ does not vary with genus and that $x(g)$
depends linearly on genus.

       Indeed, one might anticipate a distribution of the form
(\ref{ENG}) if the handles are uncorrelated.  In this case, we would posit
that handles would sprout randomly from the surface with probability
$\mu$ per unit plaquette.  This would generate the above distribution,
with $C_g \propto \mu^g/g!$, $x(g) = g$ and $\mu(g)$ independent
of $g$.  We shall refer to this behavior as the Poisson scenario.  Much
of the forthcoming analysis is devoted to a determination of whether
this scenario holds.

We start by presenting typical plots of $N_g(A)$ along with best fits to
the form (\ref{ENG})
to give a sense of
the quality of our results.

In figure 5.2 we show the behavior of genus $1$ surfaces for the SC FK
Ising clusters. Here the fit does not work.  Near the maximum, the
numerical data grows far more than the best fit allows. Genus $1$ data on
the SC lattice come indeed from fairly small surfaces (of order $100$
plaquettes, corresponding to clusters of size of tens of sites) and a
biased behavior is expected.  The situation is very different already for
genus $5$ as we show in figure 5.3.
Here the scale is given by dual surfaces of the order of $500$ plaquettes,
encompassing clusters with of order one to two hundred sites, and a
behavior closer to the continuum expectation is in order.  The fit for
figure 5.3 is indeed quite good, though some small systematic discrepancies
survive, albeit more weakly, for larger genus, where the statistical error
does eventually become very large.  On the SC lattice we find indeed, both
for the Ising model and for bond percolation, that our fits systematically
overestimate $N_g(A)$ for small $A$ and that near the peak they are
slightly too low.  Though this effect is very small already at genus $5$ it
is undoubtedly there.  We recall here that our definition of genus on the
SC lattice entails a resolution of short-distance ambiguities; perhaps this
yields a regularization that affects the geometry of moderately large
(though presumably not continuum) surfaces.  Still, the SC fits are quite
good.

We also present the fit for genus $5$ surfaces bounding
pure bond percolation clusters on the SC lattice (figure 5.4).
The results resemble those for FK clusters;
they are quite good apart from the deviations
at the peak observed previously.

The most impressive data come from measurements of Ising FK clusters
on the BCC lattice, for which there are no genus ambiguities.
Here already the genus $2$ data have an unbelievably
clean behavior (see figure 5.5).
$N_2(A)$ is peaked close to surfaces with order $250$
plaquettes, and the fit is perfect apart from the very very small
area region, where we do not expect scaling anyway.  The functional
form precisely describes both behavior for areas far below the
maximum and near the maximum itself.
Likewise, the power law plus exponential form captures all of
the relevant features of the genus $5$ data; this fit peaks at around $750$
plaquettes (see figure 5.6).
The fits continue to be superb for higher genus, though our statistics
become too poor when we reach genus $15$-$20$ to allow us to fit to the
data directly and convincingly{\footnote{Note that errors on these (and
all) plots are extremely correlated; this explains why it is possible for
our best fit to pass dead-center through so many error bars.}}.  We also
repeated these fits excluding data from surfaces of small area (less than
$100$, $150$, $200$, $250$, ..., $650$ plaquettes).  Excluding these small
areas
makes essentially no difference in the resulting fits for $g \geq 3$.

For Ising geometrical clusters on the BCC lattice, the situation is not quite
so good, at least for small genus. Indeed, for genus $2$ data
(see figure 5.7) there are large deviations from the best fit curve;
near the maximum, the fit is too low, for example.
The situation improves when we consider
higher genus data.  Genus $5$ data (see figure 5.8) agree well with
formula (\ref{ENG}). In fact, for genus larger than $4$, the fits of
(\ref{ENG}) to the data are nearly as good as the FK fits on the BCC
lattice.

In conclusion, our ansatz of equation
(\ref{ENG}) is well satisfied in the scaling limit; we will proceed now
to an analysis of the behavior of $x(g)$ and $\mu(g)$.

Let us start with $\mu(g)$ which, taking our cue from the behavior
of two--dimensional
quantum gravity\cite{GinsMoo},
we refer to as the cosmological constant. In order to
analyze our data we have used both the linear fits we have described
above and we have also computed directly the moments of the area
distribution. For data satisfying (\ref{ENG}), the cosmological
constant obeys
\beq
    \mu = \mu_{eff} \equiv \frac{\vev{A}}{(\vev{A^2} - \vev{A}^2)}
\eeq
and the exponent $x(g)$ is given by
\beq
   x(g) = x_{eff} \equiv \frac{\vev{A}^2}{(\vev{A^2} - \vev{A}^2)} -1 \ .
\eeq
Additionally, the mean area then satisfies
\beq
  \vev{A} = \frac{x_{eff}+1}{\mu_{eff}} \ .
  \protect\label{A}
\eeq
In the figure 5.9,
we show the dependence on $\mu_{eff}$ for FK clusters on the $64^3$ BCC
lattice.  The values of the cosmological constant obtained from our fits to
(\ref{ENG}) are equal (within a high degree of precision) to those obtained
from the moments for $g \geq 3$.  Clearly, the figure shows that the
cosmological constant plateaus to a constant ($0.0088 \pm 0.0002$), where
the error is mainly due to systematic, not statistical, effects. This is
one of the primary results that we present: the Ising model BCC FK
data scale with a cosmological constant which does not depend on genus and
is definitely not zero.

Figure 5.10 shows the dependence of $\mu_{eff}$ on $g$ for geometrical
clusters on the $60^3$ BCC lattice.
Here again there is clearly a plateau for the cosmological constant when $g
\geq 10$ with a value of $0.0033 \pm 0.0002$.  We also notice that for
small genus (up to genus $10$) the transient behavior of $\mu_{eff}(g)$ is
significant; this reflects deviations in the best fits of $N_g(A)$ from the
data.

Note that the value of $\mu$ essentially corresponds to the density of
handles as a function of surface area.  We find then on average that the
area needed to grow a handle is of order $110$ plaquettes on the BCC
lattice for FK clusters and $300$ plaquettes for geometrical clusters.

Next, we plot $x(g)$ (determined directly from fits) and $x_{eff}$ (from
moments) for FK clusters in the $64^3$  BCC lattice (figure 5.11).
Note that these quantities are indeed essentially identical for $g \geq 3$,
substantiating the quality of our global fits.  To see if $x(g)$ depends
linearly on $g$, we also plot the difference $x(g) - x(g-1)$ (see figure
5.12).  This difference indeed roughly appears to plateau to a
constant value, but given our statistics we cannot claim this to a great
degree of precision.  Additionally, we expect at some point that
finite-size effects will also cause deviations from linearity. From the
plateau, we would estimate the slope of $x(g)$ vs.  $g$ to be $1.25 \pm
.10$, where the quoted error is due mainly to systematic effects.

In figure 5.13, we plot $x(g)$ (determined directly from fits)
and $x_{eff}$ (from moments) for geometrical
clusters on the $60^3$ BCC lattice.
Here again these two quantities do not differ much when $g \geq 3$. Each of
these plots looks like a linear function of $g$. Figure 5.14 shows
$x_{eff}(g+1)-x_{eff}(g)$.  Again, this shows a plateau to a value constant
up to large fluctuations. In that case, the slope of $x(g)$ vs. $g$ is
$0.7\pm 0.1$.

The dependence of the mean area on genus can be measured much more
accurately (since it does not depend on a fit or on a dispersion of
moments). For FK clusters, we see from a plot of ln($\vev{A}$) vs. ln($g$)
in the small genus regime that $\vev{A}$ is not precisely linear in $g$
(see figure 5.15); in fact it scales
roughly as $g^{.85}$.
Note that such a scaling law could not hold asymptotically for large
lattices and large areas, since it would imply that surfaces could have
more handles than plaquettes.  Indeed this effective exponent slowly
increases with genus (to roughly $.90$ at $g=50$).  Thus we observe
systematic deviations (of order $15\%$) of genus dependent exponents from
their asymptotic values. From the relation ({\ref{A}}) we can conclude that
there also must be small but significant deviations from linearity of
$x(g)$ in the region $5 < g < 15$.  This suggests that the slope of $x(g)$
should decrease with greater $g$, so that the above estimate of the slope
($1.25$) may be too large.  Still our data indicates that $x(g)$ is at
least roughly linear in $g$; presumably on larger lattices with better
statistics, the systematic deviations we observe from linearity will
decrease asymptotically with large $g$.

For the geometrical case, figure 5.16 shows that the relation
between $\vev{A}$ and $g$ is nearly linear already for small genus. The
next plot (figure 5.17), showing $\ln (\vev{A})$ vs. $\ln (g)$ indicates a
small deviation for low genus.
For $2\leq g \leq 12$ we have $\vev{A} \simeq
g^{0.95}$ which becomes $g^{0.99}$ for $12\leq g \leq 24$. So we clearly
see that asymptotically we will get a linear relation between $< A >$ and
$g$ for the geometrical case.

We now return to a discussion of the data obtained for FK clusters on the
SC lattice.  Recall that generally our fits on the SC lattice have not been
nearly as good as those for data taken on the BCC lattice.  Indeed, the
results for $x$ and $\mu$ are also not nearly as clean as those obtained on
the BCC lattice, but they do substantiate our preceding qualitative
observations.  In this respect, they are important in that they allow
exhibit some degree of universality for our results.  We first show the
cosmological constant,
computed from moments, as a function of genus in figure 5.18.  Its
variation with genus is very small, being compatible with a small downward
drift superimposed on constant behavior of about $0.015$.  Thus, a handle
occurs roughly every $60$ plaquettes.  Note that we expect a larger
cosmological constant on this lattice than on the BCC lattice, since the SC
lattice contains fewer plaquettes per unit volume.

$x(g)$ also exhibits larger transient effects (due to lattice
artifacts and finite-size effects) on the SC lattice than on the
BCC lattice.   In the figure 5.19, we plot $x(g) -
x(g-1)$ for $g$ up to $15$.
This difference systematically decreases up to genus $7$ or $8$
(corresponding to significant curvature in the behavior of $x(g)$ vs. $g$
for small $g$) and then seems to level off somewhat.
In fact, at this point, the slope appears to be about $1.25$.
For
small genus, $g \simeq 15$,
we find that roughly
$\vev{A} \simeq g^{.82}$, with the exponent systematically and slowly
increasing with $g$.  Presumably, again, one would then expect that
the slope of $x(g)$ also decreases with increasing $g$.
Therefore, though the SC data is somewhat noisier
and more susceptible to lattice artifacts, we find that even the deviations
from asymptotic behavior that it exhibits are quite similar to those
measured on the BCC lattice.

Our results for $\mu$ and $x(g)$ for percolation on the SC lattice are
again quite similar.  As the figure 5.20 demonstrates
the cosmological constant does not show much variation with genus (it
is again approximately $0.015$, but
it does exhibit a small transient downward shift).  The plot of
$x(g) - x(g-1)$ in figure 5.21
resembles the one obtained in the Ising SC case, though it is even more
noisy.  We find for small genus roughly $\vev{A} \simeq g^{.81}$ with again
an exponent that increases slowly with genus.

We also examined the behavior of the constant of proportionality $C_g$ in
our fits to see if it is asymptotically compatible with the result
predicted by the Poisson scenario,
\beq C_g = C{\frac{\mu^g}{g!}}.
\protect\label{CG}
\eeq
In the figure 5.22, we plot ${\rm{ln}}(C_g) +
{\rm{ln}}(g!) - g{\rm{ln}}(\mu)$ vs.  $g$ for fits to FK cluster data on
the $L=64$ BCC lattice.
This figure indicates that $C_g$ decays more quickly than in equation
(\ref{CG}) up to about genus $10$.  Beyond that, the curve plateaus
fairly abruptly, indicating that the form of $C_g$ is indeed consistent
with the Poisson prediction above genus $10$.  We find qualitatively
identical results when we plot the same quantity extracted from FK cluster
SC lattice data and pure percolation data. Again $C_g$ decays more quickly
than Poisson indicates for small $g$ but is again compatible with the
Poisson scenario above genus $10$.

As usual, a deviation from eq. (\ref{CG}) is
expected for
low genus. Otherwise, for small $g$, $N(g)$, obtained by integrating the
area dependence
of $N_g(A)$ would behave like
\beq
N(g) \simeq C_g {\Gamma(x(g)-1) \over {\mu^{x(g)}}} \; .
\label{MCG}
\eeq
Since the slope of $x(g)$ is greater than one in this regime,
$N(g)$ would increase with
the genus if (\ref{CG}) were correct. Such an increase is
certainly not present,
thus ${\rm{ln}}(C_g)+{\rm{ln}}(g!)-g{\rm{ln}}(\mu)$ decreases initially.
For larger genus, the plateau is roughly consistent with an asymptotic
slope of $1$ for $x(g)$.

For geometrical clusters the situation is different, see figure 5.23
As above, ${\rm{ln}}(C_g)
+{\rm{ln}}(g!)-g{\rm{ln}}(\mu)$ first decreases up  to genus 4.
Then this quantity begins to grow, which seems consistent
with a slope of $x(g)$ that is less than $1$ and our observed behavior
of $N(g)$ (which we discuss next).
Here we do not really see a plateau, though the statistical errors
are very large for high genus data.

There is another ansatz which perhaps better fits the genus dependence of
$C_g$.  It was already mentioned that the deviation of $C_g$ from
(\ref{CG}) is due to the fact that $x(g)$ deviates from $g$. In the figure
5.24, we compare these deviation by displaying
${\rm{ln}}(C_g)+{\rm{ln}}(g!)-g{\rm{ln}}(\mu)$ and $(g-x(g))$ together.
On this plot,
we see that despite large statistical errors, there is an exact proportional
relation between $(g-x(g))$ and ${\rm{ln}}(C_g)+{\rm{ln}}(g!)-g{\rm{ln}}
(\mu)$. This indicates
that (\ref{CG}) could be changed to
\beq
C_g=C (e^{\beta})^{g-x(g)} {\mu^g \over g!}
\protect\label{newcg}
\eeq
with $\beta$ the constant of proportionality (which is close to $10$ in our
case.)
Of course, the above relation reduces to the Poisson prediction
when the slope of $x(g)$ is $1$.

By inserting (\ref{newcg}) in (\ref{ENG}), we have :
\beq
N_g(A)= C {(e^\beta \mu)^g \over g!} ({A\over e^\beta})^{x(g)} e^{-\mu(g) A}
\eeq
If we now redefine $m=e^\beta \mu$ and $a={A\over e^\beta}$ then this modified
ansatz can be expressed simply as
\beq
N_g(A)= C {m^g \over g!} a^{x(g)} e^{-m a }.
\eeq
This ansatz reduces to the Poisson prediction only
when the slope of $x(g)$ is $1$.
If we assume the above form (\ref{newcg})
for $C_g$ together with a linear dependence of $x(g)$ with
slope not equal to $1$, then the sum of $N_g(A)$ over $g$ (which
converges rapidly)
will not asymptotically behave as a power law in $A$.  This contradicts
our earlier expectations and observations, based on the scaling behavior
of $N(V)$ and $V \sim A$.

  The Poisson scenario provides us with one further related prediction.  It
implies that asymptotically the number of surfaces of genus $g$, $N(g)$,
should be proportional to
$g^{-\tau}$.   For the modified ansatz (\ref{newcg}), $N(g)$ will
only exhibit asymptotic power law behavior if and only if the slope of
$x(g)$ is $1$.
In the next four figures (figures 5.25-5.28),
we show log-log plots of the genus dependence of $N(g)$
for FK clusters and geometrical clusters on the
BCC lattice and FK
clusters and pure bond percolation on the SC lattice.
In all four cases, these plots appear to be quite linear.  Our fits for FK
clusters on both the BCC and SC lattices yield a scaling exponent of
$2.00 \pm 0.01$ in the region $6 \leq g \leq 24$.  For the geometrical
clusters, the scaling exponent is $2.02 \pm 0.01$ in the same region.

In this case, the results for percolation are a bit different.
We observe a systematic upward drift in the exponent for low genus.
For instance, on a window of $6 \leq g \leq 12$, we obtain an exponent
of $1.90 \pm 0.01$ (as compared to $2.00 \pm 0.01$ for BCC FK,
$1.99 \pm 0.01$ for SC FK clusters and $1.99\pm 0.01$ for BCC geometrical
clusters).  The exponent is closer to the Ising
exponent on higher genus windows, albeit with a large statistical
error.  For example, we obtain for percolation $1.98 \pm 0.03$ in the
window $12 \leq g \leq 18$ (compared to $1.98 \pm 0.02$ for BCC
Ising, $1.98 \pm 0.05$ for SC Ising and $2.06 \pm 0.01$ for BCC geometrical
clusters) and $1.97 \pm 0.05$ in the
window $18 \leq g \leq 24$ (compared to $1.98 \pm 0.07$ for BCC
Ising, $2.03 \pm 0.09$ for SC Ising and $2.02\pm 0.02$ for BCC geometrical
clusters).

We definitely do observe power law behavior (as predicted by Poisson)
but our exponents consistently
are roughly $10$ percent lower than $\tau$ (except in the case of geometrical
clusters where the difference is only $5$ percent); only in the case of
bond percolation do we see any asymptotic upward drift in this exponent.
Yet, given our experience with measuring other exponents in these
systems, it seems reasonable that this discrepancy from Poisson could
be attributed to systematic effects.

     In conclusion, our genus data indicate that all scaling clusters
examined satisfy the ansatz (\ref{ENG}) with a nearly constant
$\mu$ and an an exponent $x$ that asymptotically appears to
depend linearly on $g$.  We do, though, observe variations
in the slope of $x$ from $1$ and other deviations (in the behavior
of the overall coefficient $C_g$, e.g.) from the Poisson scenario.
It is unclear, though, whether these deviations are significant;
they appear to be somewhat inconsistent with other observations.
We also know that  finite-volume effects can in some cases induce systematic
deviations in our exponents of at least 15-20 percent.  Still, for geometrical
clusters, we do not directly observe large finite volume effects
in the measurement of $x(g)$; its value is rather stable as $L$ changes
from $60$ to $100$.  Perhaps larger scale simulations are needed to
properly determine the asymptotic form of $x(g)$.

\subsection{Loop Scaling and Blocked Spins}

One might wonder if there is any characteristic of the geometrical
clusters that reflects the Ising phase transition at $T_c$,
rather than the percolation transition at $T_p$.
This cannot be a simple extensive property of the surfaces
such as their total area or topology, as we have seen.
Rather one needs a finer measure of their distribution, which in
particular properly reflects surface roughness.
We have found one such measure by taking
cross-sections of the surfaces.
Consider the ensemble of loops formed by the intersection of
the set of cluster boundaries with an arbitrary two-dimensional plane.
We have found that the
distribution of lengths of these loops is sensitive to the
critical dynamics of the Ising fixed point.

To begin, recall the three-dimensional structure
of boundaries of geometrical clusters as $T$ is increased beyond $T_p$,
particularly to $T=T_c$.  For $T > T_p$, two percolated
clusters of opposite sign span the lattice.  For $T$ not so close
to $T_c$, we expect that the characteristics of the Ising interaction
will not influence the large--scale structure of these percolating clusters.
The percolating clusters (assuming the transition at $T_p$ is indeed
in the universality class of pure percolation) should then be described
by the `links, nodes and blobs' picture developed for the infinite
clusters of pure percolation in dimensions below $d_c=6$
\cite{Stauffer,DeGennes}\index{Links, nodes and blobs}.
In this description, the links form the
thin backbones of the cluster; they are connected together at the nodes
which occur roughly every percolation correlation
length $\xi$.  Most of the volume
of the cluster lies in dangling ends emanating from the backbones.
The backbones are not simply-connected.  Rather, they contain
fingers which fuse together to generate the handles
that we measure, thus forming blobs with diameter
up to size $\xi$.

     A cross section of the boundaries of these networks
of tangled thin tubes would presumably be composed of a set of
small lattice--sized loops.  To check this, we examined the phase
boundaries between up and down spins on planar slices of both
the SC and BCC lattices.  In figure 5.29, we show a
log--log plot of $N(l)$, the number of loops of length $l$, versus
$l$ taken at the percolation temperature $
\beta_p = .232$ on the SC lattice.  The
curve exhibits a sharp drop--off, indicating indeed
that these slices contain only small loops.
As we dial the temperature up towards $T_c$, we find that
larger loops begin to appear in the slices.
In figure~5.30, we present a `movie' of four
consecutive slices at $T_c$.  Loops that are small, large and
intermediate sized are present in each of these slices.
In fact,
at $T_c$, we find loops at all scales; $N(l) \sim l^{-\tau '}$!
This scaling is depicted in the log--log plot in
figure 5.31.
As in figure 5.1, we observe a small bump at the end
of the distribution followed by a rapid drop--off.  These deviations
from scaling are again due to the influence of the finite--size
of the lattice on the largest loops.
All of the largest loops must bound the
two percolating clusters, since there are no intermediate size
geometrical clusters at $T_c$.
The loops themselves have a non--trivial fractal structure;
we determined that the number of sites enclosed within a loop of length $l$
scales as $A(l) \sim l^{\delta '}$.

 From these measurements, we estimated that $\tau ' = 2.06(3)$ and
$\delta ' = 1.20(1)$.  These values are probably not very accurate,
however.  As in the determination of $\tau$ from the
behavior of $N(V)$, corrections to scaling and finite--size effects
are a source of large systematic errors.  These systematic
effects were only of order $1-2\%$ for $\delta$; thus we suspect that
our estimate of $\delta '$ is considerably better
than that of $\tau '$.  Carrying out these measurements also
required a resolution of certain ambiguities.  In particular, since
the boundaries of domains
self--intersect on slices of the cubic lattice, we had
to pick a prescription (effectively another short--distance regularization)
to define loops.  Additionally, the enclosed area is not
well--defined for loops
that wind around the (periodic) lattice.  We thus chose
to exclude loops with non--zero winding number from
consideration.  Also, we note that these measured values presumably
suffer from large systematic corrections because they do not
satisfy the
relation $\tau ' = 1 + \delta '$, which
can be derived through scaling arguments \footnote{ See section
\ref{2dresults}}.
This relation also holds for the corresponding indices that
describe the distribution of self--avoiding loops that bound
clusters in the 2--$d$ Ising model at the Curie temperature.  In that
case, $\tau ' \sim 2.45$. Finally, we found that the scaling
behavior of loops on slices
slowly disappeared as we
continued to increase the Ising temperature.  At
$\beta = .18$ on $L=150$ SC lattices, we observed that very large loops
were again exponentially suppressed in the distribution $N(l)$.

Should we be surprised by the presence of this `loop scaling' at
$T_c$?  The following argument, due to Antonio Coniglio, indicates that
this result is at least plausible \cite{priv2}.  First, note that in the  $T
\rightarrow \infty$ limit, the distribution of loops and geometrical
clusters is that of pure site percolation with $p = .5$.  For site
percolation on the
square lattice, $p_c \sim .59$ so that if only half the sites
contain identical spins, then the distribution of loops
and clusters should be
governed by a finite correlation length. Now consider turning on
the Ising couplings in the $x$ and $y$ directions.  As the spins
become correlated, the critical concentration\footnote{Note
that we can adjust the relative concentration of up and down spins
by also adding a magnetic field.} needed for percolation should
diminish.  At the Curie temperature for the 2--$d$ Ising model
($T_c^{d=2}$) this critical concentration decreases to $.5$
and geometrical clusters and their boundaries percolate.
In two dimensions, this critical concentration cannot be less
than $.5$, since generically two percolating clusters cannot
span a single lattice \cite{co1}.   Imagine next turning
on the Ising coupling in the $z$ direction while tuning
the $x$ and $y$ couplings to remain at criticality.  If
the critical concentration remains $.5$ as the system reaches
the 3--$d$ Curie temperature, then one would find a scaling
distribution of clusters and boundaries on 2--$d$ slices.
On the other hand, we cannot rule out the possibility that the
critical concentration again increases above $.5$; then
we would never expect to find scaling of loops on slices of
the 3--$d$ Ising model.

      We also observed scaling behavior of loops on the BCC lattice.
In particular, only small loops were found at $T_p$ while scaling
of $N(l)$
with the values $\tau ' = 2.23(1)$ and $\delta ' = 1.23(1)$
occurred at $T_c$.  The uncertainty in the value of $T_c$ probably
leads to a significant systematic error in the estimate of these
exponents.  They do obey the anticipated relation $\tau ' = 1 +
\delta '$; $\delta '$ is not particularly far from the estimate
extracted from the SC data.  Note that on slices of the BCC lattice, which
are triangular, there is no longer any ambiguity in the definition
of loops.
In this case, $N(l)$ apparently satisfies a power--law
distribution, with a temperature--dependent exponent, for all $T>T_c$!
This observation can be fully understood theoretically, since
the percolation threshold
on triangulated lattices equals $.5$.  Therefore, we definitely
expect to observe loop scaling at $T =
\infty$ with scaling exponents characteristic of 2--$d$ percolation
($\tau ' \sim 2.05$ and $\delta ' = 1$).  Since lowering the temperature
increases correlations between spins, we expect to find percolated
clusters on slices for all $T$.  For $T<T_c$, however, minority
spins cannot percolate on 2--$d$ slices because, as stated above,
only one infinite cluster can span a lattice.  Thus the minority
spins and the loops that enclose them must percolate at $T_c$
on 2--$d$ slices of the 3--$d$ Ising
model on the BCC lattice.  If we assume that this phenomenon is
independent of the particular lattice type, then it follows that
loop scaling should always occur at $T_c$.  A similar situation occurs
for the 2--$d$ Ising model on the triangular lattice:  one can
argue that the distribution
$N(l)$ again scales as a power law for all $T > T_c$ because $p_c = 1/2$
on triangulated lattices.

It also seems reasonable that the presence of loop scaling
may be related to the vanishing of the surface tension of
the Ising model at $T_c$.  Antiperiodic boundary conditions in one
direction (say $\hat z$) force the appearance of an interface
transverse to $\hat z$.
The surface tension vanishes when the free energy
of a system with such anti-periodic boundary conditions
equals the free energy of a system with periodic boundary conditions
in $z$\index{Ising model!surface tension}.
Consider a slice through the lattice in the $x$-$z$ plane; it cuts
the interface along a loop that winds across the $x$-direction.
Vanishing surface tension allows this loop to wander freely
due to the unsuppressed surface fluctuations.  Thus one expects
to find that the probability distribution for this loop to
have length $l$ is not cut off at large $l$.  Furthermore,
the probability to find a loop of length $l$ much larger than the linear
size of the system $L$ should not care whether the loop is
topologically wound across $x$.
Hence, vanishing surface tension and loop scaling should be related
phenomena.

We now comment on the significance of this scaling.
  As we noted in the previous two sub--sections, the geometrical
cluster boundaries do not in the least resemble surfaces (in
the continuum limit) at $T_p$.  The presence of large loops at
$T_c$ might indicate that the boundaries grow large long handles.
A visual examination of successive slices qualitatively indicates
that this is not so.  Large loops seemingly always vanish after
several consecutive slices.  Indeed, it is difficult to envision
a smooth surface that decomposes into a scaling distribution of loops
along arbitrary slices.

  It should also be noted that the exponent $\tau '$
is probably not directly related to the magnetic or thermal exponents
of the 3--$d$ Ising model.  More generally, it may not
be associated with the behavior
of correlation functions of local operators in a unitary
quantum field theory.
This is true also for loops bounding
clusters in the 2--$d$ Ising model.  For in all of these cases,
the scaling of geometrical clusters
is determined by the geometric effects associated with percolation as
well as the long--range correlations due to Ising criticality.
Still, this scaling law describes physics that in
principle is observable, perhaps by counting domains in sections of
crystals that lie in the universality class of 3--$d$ Ising.
It would thus be quite interesting to construct a theoretical
scheme to compute (approximately) the value of $\tau '$.
These loops are significantly `rougher' than the corresponding
boundaries in the 2--$d$ Ising model, since the exponent $
\delta ' $ is lower here.  They gain more kinetic
energy because they are given an extra dimension in which to
vibrate; perhaps this is responsible for their increased roughness.

Ideally, we would like to
view these loops as string states that evolve in Euclidean
time (perpendicular to the slices).  Their dynamics
is described by the transfer matrix determined from Boltzmann
factors associated with their creation, destruction, merging and
splitting\index{Transfer matrix}.
We have thus found that the ground state wave functional (string field)
of this transfer matrix is peaked around
configurations that describe a scaling distribution of loops.
These loops seemingly bear little relation to free strings,
though, because they interact strongly by splitting and joining
every few lattice spacings \footnote{In practice, this makes an
analysis of the transfer matrix a formidable task.}.
This is why the entire history of the loop ensemble largely consists of
a single surface, whose gross properties have little to do with
the critical dynamics.
One might hope that some sort of perturbative string
description could still be viable if the strength of this interaction
were just a short--distance artifact; i.e. if the string coupling
diminished towards zero in the infrared.  To gauge whether this
is likely, we blocked spins in our simulations to measure the renormalization
group flow of the operator that couples to the total Euler character
summed over all cluster boundaries.  In particular, during simulations
on L=128 SC and BCC lattices, we blocked spins, using the majority
rule and letting our random number generator decide
ties\index{Renormalization group!Blocked spins}.  At
each blocking level, we reconstructed clusters and boundaries and
then measured the genus summed over surfaces.  We present the results
of this analysis in table \ref{bstable}; data was taken at $\beta_c =
.221651$ on the SC lattice and $\beta_c = .0858$ on the BCC lattice.

The results are not so conclusive.  In particular,
since we lack a very precise determination of the Curie temperature on the
BCC lattice, it is likely that by the final blocking the couplings have
flowed significantly into either the high or low--temperature regimes.
Thus, one should probably not take the increase in genus density
in the final two blockings on the BCC lattice seriously.  This effect
is not a problem on the SC lattice, where we fortunately know the
critical temperature (based on previous Monte Carlo Renormalization
Group measurements) to very high accuracy.  On the other hand, we
suspect that the small $L$ blocked values on the SC lattice may be
unreliable, due to ambiguity in the definition of genus.  We can at
least infer that the genus density decreases a bit during the first
few blockings, indicating that the coupling $\exp (-\kappa)$
does at least slowly diminish at the beginning of the RG flow.  There is
no clear indication, however, that the flow continues on to the weak
string coupling regime.  In fact, we would naively expect that
this `genus' operator is irrelevant, since it involves couplings between
next--nearest neighbor spins.  Hence, we would not anticipate that the genus
would decrease dramatically upon blocking.
One might also object to our choice of blocking scheme.
Indeed, perhaps it might be more appropriate to somehow block
the cluster boundaries themselves rather than the spins.
In practice this
would probably be technically difficult.

\section{Assessment}

     The prospects for passing from the Curie point to the regime
in which surfaces are weakly coupled are addressed in the work of
Huse and Leibler \cite{HuseLei}.  They qualitatively
map out the phase diagram of
a model of self--avoiding surfaces with action (\ref{action}).
The large $\kappa$ (large coupling to total Euler character) regime
of their model lies in a droplet crystal phase, where the large
percolated surface has shattered into a lattice of small disconnected
spheres.  Such a configuration maximizes the Euler density; it
clearly does not correspond to a theory of surfaces.  By
estimating the free energy difference between phases, they argue
that the transition to this droplet crystal is first order.  Given
this picture, there seems to be little evidence for the existence of
a fixed point describing a weakly coupled theory of surfaces
near the Curie point of the Ising model.  Nevertheless, we cannot
definitely exclude the possibility that there is still some path
which we have not considered to a weak--coupling theory.

 In conclusion, it appears that evidence of a continuum theory
of surfaces has eluded us in our investigation of Ising cluster
boundaries.   We have found, however, that these cluster boundaries
do exhibit an intriguing fractal structure that does not typically
appear in models of lattice surfaces.

\section{Acknowledgments}
We would like to thank Stephen Shenker for essential discussions which led
to our investigations.  We also greatly benefitted from discussions and
correspondence with Mark Bowick, Antonio Coniglio, Francois David, Bertrand
Duplantier, John Marko, Pierre Pujol and Jim Sethna. We are also grateful
to NPAC for their crucial support.  This work was supported in part by the
Dept. of Energy grants DEFG02-90ER-40560, DEFG02-85ER-40231, the
Mathematical Disciplines Institute of the Univ. of Chicago, funds from
Syracuse Univ., by the Centre National de la Recherche Scientifique, by
INFN and the EC Science grant SC1*0394.

\section{Appendix}

     We now discuss the derivation of the scaling relation \ref{L4},
which asserts that  $\tau  = \delta + 1$.  These arguments are meant
to be descriptive, not mathematically rigorous.  Consider a configuration
of loops bounding clusters on a  slice $\cal{D}$ of size
$R \times R$.  For a visualization of these loops, see Fig.~5.30.
As before, the mean area $A(l)$ within a loop of size
$l$ scales as $l^{\delta}$; the number of loops of size $l$, $N(l)$, is
proportional to $R^2 l^{-\tau}$.  These formulas are at least valid in the
regime $1 \ll l \ll R^{2/\delta}$.

For $\lambda$ slightly larger than $1$, consider the quantity
\beq
\label{appeq}
F({\bar l},\lambda) \equiv  \int_{\bar l}^{\lambda {\bar l}}N(l)A(l)dl
\propto R^2 {\bar l}^{\gamma} \frac{\lambda ^{\gamma} - 1}{\gamma};
\eeq
$\gamma \equiv \delta + 1 - \tau$.  To a first approximation, $F$ measures
the area enclosed in loops of size between ${\bar l}$ and $\lambda {\bar l}$.
This correspondence would not be exact if, for instance, one of the loops of
size ${\bar l}$ were embedded
in a loop of size $\lambda{\bar l}$.  The loops are self--avoiding and tend
to meander through the slice ${\cal D}$, so such an embedding is highly
unlikely for $\lambda$ close to $1$.  In this case, the over--counting due
to these embeddings is negligible and $F$ cannot be much greater than $R^2$.
It follows then that $\gamma$ cannot be greater than zero, for otherwise
the contribution from ${\bar l}^{\gamma}$ would over--saturate this limit
for large ${\bar l}$.

We now argue that likewise $\gamma $ cannot be negative.  Consider a
fixed value of $\lambda$ not necessarily very close to unity; e.g.
$\lambda = 10$.  If $\gamma$
were negative, then for large ${\bar l}$, the area enclosed within loops
with ${\bar l} < l < \lambda {\bar l}$ would be a negligible fraction of
the area of the entire slice.  It follows from the self--similarity of
the percolation clusters that this cannot be so.  Let us take an arbitrary
subdomain of extent ${\bar l}^{2/\delta} \times {\bar l}^{2/\delta}$.
By self--similarity, there should generically exist a cluster that barely
spans this subdomain; its surrounding loop should have size of order
${\bar l}$ and will enclose a non--negligible portion of this subdomain
\footnote{As is apparent from Fig.~5.30, the loops are fat.  Clearly
in their meanderings, they will cut off and surround large islands in the
regions of the slice that they traverse.}.
We can tile the slice ${\cal D}$ with these subdomains; loops of size of
order ${\bar l}$ (${\bar l} < l < \lambda {\bar l}$) will
then cover a significant fraction of the entire slice.
We thus conclude that $\gamma = 0$ and this scaling relation holds.

\vfill
\newpage
\topmargin -.3in
\flushbottom
\begin {flushleft}
{\bf Figure Captions}
\end{flushleft}
\begin{itemize}
\item[Fig.~3.1] The Wigner--Seitz cell of the BCC lattice with next--nearest
neighbor interactions.

\item[Fig.~4.1] $\ln N(L)$ vs. $\ln L$ for geometrical clusters
on the $1000\times 1000$ triangular lattice.

\item[Fig.~4.2] $\ln A(L)$ vs. $\ln L$ for geometrical clusters
on the $1000\times 1000$ triangular lattice.

\item[Fig.~5.1] $\ln N(V)$ vs. $\ln V$ for FK clusters
on the $L=64$ SC lattice.

\item[Fig.~5.2] The number of genus $1$ surfaces at $T_c$
as a function of dual surface area $A$ for FK clusters on the $L=64$ SC
lattice, with a best fit to the functional form given in equation
(\ref{ENG}).

\item[Fig.~5.3] As in the previous figure, but for genus 5.

\item[Fig.~5.4] As in the previous figure, but for $3d$ bond percolation
clusters.

\item[Fig.~5.5] As in the previous figure, but for FK clusters on the
$L=64$ BCC lattice and for genus $2$.

\item[Fig.~5.6] As in the previous figure, but for genus $5$.

\item[Fig.~5.7] The number of genus $2$ surfaces at $T_p$ as a function of
dual surface area $A$ bounding minority (geometrical) clusters on the
$L=60$ BCC lattice.

\item[Fig.~5.8] As in the previous figure, but for genus $5$.

\item[Fig.~5.9] The dependence of  $\mu$ (extracted from the
moments of the area distribution) on genus for FK clusters
on the $L=64$ BCC lattice at $T_c$.

\item[Fig.~5.10] The dependence of $\mu$ (extracted
from moments) on genus for surfaces bounding minority (geometrical)
clusters on the $L=60$ BCC lattice at $T_p$.

\item[Fig.~5.11] The dependence of $x$ (extracted from
direct fits to (\ref{ENG})
and moments) on genus for FK clusters on the $L=64$ BCC lattice at $T_c$.

\item[Fig.~5.12] The dependence of $x(g) - x(g-1)$ on genus for FK clusters
on the $L=64$ BCC lattice at $T_c$.

\item[Fig.~5.13] The dependence of $x$ (extracted from
direct fits to (\ref{ENG}) and moments) on genus for surfaces bounding
minority (geometrical) clusters on the $L=60$ BCC lattice at $T_p$.

\item[Fig.~5.14] The dependence of $x(g) - x(g-1)$ on genus for surfaces
bounding minority (geometrical) on the $L=60$ BCC lattice at $T_p$.

\item[Fig.~5.15] The dependence of ln($\vev{A}$) on ln(g) for FK clusters
on the $L=64$ BCC lattice at $T_c$.

\item[Fig.~5.16] The dependence of $\vev{A}$ on genus for surfaces
bounding minority (geometrical) on the $L=60$ BCC lattice at $T_p$.

\item[Fig.~5.17] The dependence of ln($\vev{A}$) on ln(g) for
surfaces bounding minority (geometrical) on the $L=60$ BCC lattice at
$T_p$.

\item[Fig.~5.18] The dependence of  $\mu$ (extracted from the
moments of the area distribution) on genus for FK clusters
on the $L=64$ SC lattice at $T_c$.

\item[Fig.~5.19] The dependence of $x(g) - x(g-1)$ on genus for FK clusters
on the $L=64$ SC lattice at $T_c$.

\item[Fig.~5.20] The dependence of  $\mu$ (extracted from the
moments of the area distribution) on genus for $3d$ bond percolation
clusters on the $L=64$ SC lattice.

\item[Fig.~5.21] The dependence of $x(g) - x(g-1)$ on genus for $3d$ bond
percolation clusters on the $L=64$ SC lattice.

\item[Fig.~5.22] The dependence of ${\rm{ln}}(C_g) +
{\rm{ln}}(g!) - g{\rm{ln}}(\mu)$ on genus for FK clusters
on the $L=64$ BCC lattice at $T_c$.

\item[Fig.~5.23] As in the previous figure, but for surfaces bounding
minority (geometrical) clusters on the $L=60$ BCC lattice at $T_p$.

\item[Fig.~5.24] The dependence of ${\rm{ln}}(C_g) +
{\rm{ln}}(g!) - g{\rm{ln}}(\mu) $ ($^{\framebox[2mm]{$\;$}})$ and
$10*(g-x(g)) (\bullet$) on
genus for surfaces bounding
minority (geometrical) clusters on the $L=60$ BCC lattice at $T_p$.

\item[Fig.~5.25] The dependence of ln($N(g)$) on ln(g) for FK clusters
on the $L=64$ BCC lattice at $T_c$.

\item[Fig.~5.26] As in the previous figure, but for surfaces bounding
minority (geometrical) clusters on the $L=60$ BCC lattice at $T_p$.

\item[Fig.~5.27] As in the previous figure, but for FK clusters on the
$L=64$ SC lattice at $T_c$.

\item[Fig.~5.28] As in the previous figure, but for $3d$ bond percolation
clusters on the $L=64$ SC lattice.

\item[Fig.~5.29] A log--log plot of the distribution of loops
of length $l$ on slices of an $L=60$ SC lattice at $T_p$.

\item[Fig.~5.30]  Four consecutive slices of a representative configuration
of geometrical clusters at $T_c$.

\item[Fig.~5.31] A log--log plot of the distribution of loops of
length $l$ on slices of an $L=150$ SC lattice at $T_c$.
\end{itemize}

\vfill
\newpage
\topmargin -.3in
\flushbottom
\begin {flushleft}
{\bf Tables}
\end{flushleft}
(FKC = FK clusters, GC = Geometrical clusters,
BP = $3d$ Bond percolation.)
\begin{table}[h]
\begin{tabular}{|l||l|l|l|l|l|l|l|l|} \hline
cluster type & lattice & size & no. of sweeps & $\beta$ \\
\hline
GC &square & 500 & 100000 & 0.44068 \\
GC & square & 1000 & 25000 & 0.44068 \\
GC &triangular & 500 & 100000 & 0.27465 \\
GC &triangular & 1000 & 25000 & 0.27465 \\
GC &triangular & 500 & 100000 & 0 \\
GC &triangular & 1000 & 25000 & 0 \\
\hline
\end{tabular}
\protect\caption[CT_ONE]{A record of the number of sweeps performed
on two-dimensional lattices.
\protect\label{T_ONE}}
\end{table}
\begin{table}[h]
\begin{tabular}{|l||l|l|l|l|l|l|l|l|} \hline
cluster type & lattice & size & no. of sweeps & $\beta$ \\
\hline
FKC & SC & 32 & 6000000 & 0.221651 \\
FKC & SC & 64 & 250000 & 0.221651 \\
FKC & BCC & 64 & 300000 & 0.0858 \\
GC & BCC & 30 & 500000 & 0.0959 \\
GC & BCC & 60 & 500000 & 0.0959 \\
GC & BCC & 100 & 50000 & 0.0959 \\
BP & SC & 32 & 50000 & -- \\
BP & SC & 64 & 11000 & -- \\
\hline
\end{tabular}
\protect\caption[CT_TWO]{A record of the number of sweeps performed
on three-dimensional lattices.
\protect\label{T_TWO}}
\end{table}
\begin{table}[h]
\begin{tabular}{|l||l|l|l|l|l|l|l|l|} \hline
cluster type & lattice & size & no. of sweeps & $\beta$ \\
\hline
GC & SC & 60 & 40000 & 0.2216 \\
GC & SC & 150 & 4000 & 0.2216 \\
GC & BCC & 150 & 1000 & 0.0858 \\
\hline
\end{tabular}
\protect\caption[CT_THREE]{A record of the number of three-dimensional
configurations produced for an analysis of cross-sectional slices.
\protect\label{T_THREE}}
\end{table}
\begin{table}[h]
\begin{tabular}{|l||l|l||l|l|l|l|l|l|} \hline
interval of $l$  & $\tau_{500}$ & $\delta_{500}$ & $\tau_{1000}$ &
$\delta_{1000}$\\
\hline
\, 40 --- 2500  &  2.389  & 1.444 &  2.421  & 1.450\\
100 --- 2500  &  2.382  & 1.445 &  2.419  & 1.451 \\
\, 40 --- 1500  &  2.403  & 1.444 &  2.421  & 1.448 \\
100 --- 1500  &  2.396  & 1.445 &  2.416  & 1.450 \\
\hline
\end{tabular}
\protect\caption[CT_ONE2]{Square lattice, $500 \times 500$ and $1000 \times
1000$.
\protect\label{T_ONE2}}
\end{table}
\begin{table}[h]
\begin{tabular}{|l||l|l||l|l|l|l|l|l|} \hline
interval of $l$  & $\tau_{500}$ & $\delta_{500}$ & $\tau_{1000}$ &
$\delta_{1000}$ \\
\hline
\, 20 --- 2500  &  2.431  & 1.454 &  2.440  & 1.452 \\
100 --- 2500  &  2.427  & 1.455 &  2.436  & 1.452 \\
\, 20 --- 1500  &  2.438  & 1.454 &  2.444  & 1.454 \\
100 --- 1500  &  2.433  & 1.455 &  2.439  & 1.455 \\
\hline
\end{tabular}
\protect\caption[CT_TWO2]{Triangular lattice $500 \times 500$ and $1000
\times 1000$.
\protect\label{T_TWO2}}
\end{table}
\begin{table}[h]
\begin{tabular}{|l||l|l||l||l|l|l|l|l|} \hline
interval of $l$  & $r^c_{l,500}$ & $ r^c_{l,1000}$ & interval of $l$  &
$r^\infty_{l,500}$ & $r^\infty_{l,1000}$\\
\hline
100 --- 800  &  2.471  & 2.471 & 200 --- 800  &  2.219  & 2.218 \\
200 --- 800  &  2.472  & 2.471 & 300 --- 800  & 2.218 & 2.217 \\
 100 --- 1200  &  2.472  & 2.471 & 200  --- 1200  & 2.220  & 2.218 \\
200 --- 1200  &  2.473  & 2.471 & 300 --- 1200  &  2.220  & 2.218 \\
\hline
\end{tabular}
\protect\caption[CT_ONE6]{$r_l$ for the $500 \times 500$ and $1000 \times
1000$ triangular lattices. }
\end{table}
\begin{table}[h]
\begin{tabular}{|l||l|l|l|l|l|} \hline
 lattice & 128 & 64 & 32 & 16 & 8 \\
\hline
BCC & .049 (3) & .039 (3) & .037 (3) & .039 (3) & .044 (3) \\
\hline
SC  & .021 (2) & .020 (2) & .018 (2) & .015 (2) & .012 (1) \\
\hline
\end{tabular}
\protect\caption{\label{bstable}The mean genus per lattice site
at $T_c$ for blockings ($L=8,16,32$ and $64$) of an $L=128$ lattice.}
\end{table}


\newpage
\begin{thebibliography}{99}
\def\cmp{{\it Commun. Math. Phys.}}
\def\np{{\it Nucl. Phys.}}
\def\pl{{\it Phys. Lett.}}
\def\prl{Phys. Rev. Lett.}
\def\pr{Phys. Rev.}
\def\pre{{\it Phys. Rep.}}
\def\mpl{{\it Mod. Phys. Lett.}}
\def\jpa{J. Phys. A}
\bibitem{string}
A.~Polyakov, Phys. Lett. B 82 (1979) 247;
E.~Fradkin, M.~Srednicki and L.~Susskind, Phys. Rev. D 21
(1980) 2885;
C.~Itzykson, Nucl. Phys. B 210 (1982) 477;
A.~Casher, D.~F\oe rster and P.~Windey, Nucl. Phys. B 251 (1985) 29;
Vl.~Dotsenko and A.~Polyakov, {\it in} Advanced Studies in Pure
Math. 15 (1987).

\bibitem{clt1}
E.~Br\'{e}zin and V.A.~Kazakov, Phys. Lett. B 236 (1990) 144;
M.R.~Douglas and S.H.~Shenker, Nucl. Phys. B 335 (1990) 635;
D.~J.~Gross and A.~A.~Migdal, Phys. Rev. Lett. 64 (1990) 127.
\bibitem{tachyon}
G.~Parisi, in Proceedings of the Third Workshop on
Current Problems in High Energy Particle Physics, John Hopkins Conference,
Florence 1979;
G.~Parisi, J.-M.~Drouffe and N.~Sourlas, Nucl. Phys. B 161 (1979) 397;
B.~Durhuus, J.~Frohlich and T.~Jonsson,  Nucl. Phys. B 240 (1984) 453;
J.~Ambjorn, B.~Durhuus, J.~Frohlich and P.~Orland, Nucl. Phys. B 270
(1986) 457;
M.~E.~Cates, Europhys. Lett. 8 (1988) 719.

\bibitem{david2}
F.~David, Europhys. Lett. 9 (1989) 575.

\bibitem{HuseLei}
D.~Huse and S.~Leibler, J. de Physique 49 (1988) 605.

\bibitem{previous}
M.~Karowski and H.~J.~Thun, Phys. Rev. Lett. 54 (1985) 2556;
R.~Schrader, J. Stat. Phys. 40 (1985) 533.

\bibitem{Gliozzi}
M.~Caselle, F.~Gliozzi and S.~Vinti, Turin Univ. preprint DFTT--12--93;
Nucl. Phys. Proc. Suppl. B 34 (1994) 726.

\bibitem{David}
F.~David, Jerusalem Gravity (1990) 80.

\bibitem{Stauffer}
D.~Stauffer and A.~Aharony, Introduction to Percolation Theory
(Taylor and Francis, London 1992).

\bibitem{CambNaue}
J.~Cambier and M.~Nauenberg, Phys. Rev. B 34 (1986) 8071.

\bibitem{FK}
C.~M.~Fortuin and P.~W.~Kasteleyn, Physica 57 (1972) 536.

\bibitem{CK}
A.~Coniglio and W.~Klein, J. Phys. A 13 (1980) 2775.

\bibitem{Sokal}
R.G.~Edwards and A.D.~Sokal, Phys. Rev. D 38 (1988) 2009;
A.D.~Sokal, in Monte Carlo methods in statistical mechanics:
Foundations and algorithms, NY preprints based on lectures at the
Troisi\`eme Cycle de la Physique en Suisse Romande, June 1989.

\bibitem{Hu}
C.~-K.~Hu, \pr B 29 (1984) 5103.

\bibitem{Wang}
J.-S.~Wang, Physica A 161 (1989) 149.

\bibitem{a1} B.~Duplantier,  J. Stat. Phys. 49 (1987) 411;
                           Physica D 38 (1989) 71;
  H.~Saleur and B.~Duplantier, Phys. Rev. Lett. 58 (1987) 2325.

\bibitem{cardy} J.~Cardy, Lecture notes of Les Houches Summer School, 1994,\\
cond-mat@babbage.sissa.it \# 9409094

\bibitem{SW}
R.~H.~Swendsen and J.-S.~Wang, Phys. Rev. Lett. 58 (1987) 86.

\bibitem{sw}
R.~H.~Swendsen and J.-S.~Wang, \jpa 13 (1980) 2775.

\bibitem{Hasenbuch}
M.~Hasenbusch and K.~Pinn, Munster Univ. preprint MS--TIP--92--24.

\bibitem{kirk}
S.~Kirkpatrick, in Ill-Condensed Matter, ed. by R.~Balian,
R.~Maynard and G.~Toulouse (North-Holland, Amsterdam 1979) p.321.

\bibitem{a2} B.~Duplantier, private communication, unpublished

\bibitem{a3} J.~L.~Cambier and M.~Nauenberg, Phys. Rev. B 34
(1986) 8071

\bibitem{a4} C.~Vanderzande and A.~L.~Stella,
J. Phys. A 22 (1989) L445


\bibitem{ourletter}
V.~Dotsenko, G.~Harris, E.~Marinari, E.~Martinec, M.~Picco and P.~Windey,
Phys. Rev. Lett. 71 (1993) 811.

\bibitem{proceedings}
V.~Dotsenko, G.~Harris, E.~Marinari, E.~Martinec, M.~Picco and P.~Windey,
Proccedings of 1993 Cargese Workshop, to appear, hep-th/9401129.

\bibitem{Bake}
G.~Baker,B.~Nickel,M.~Green and D.~Meiron, Phys. Rev. Lett. 36
(1976) 1351.

\bibitem{DombZinn2}
C.~Domb, in Phase Transitions and Critical Phenonema, vol.~3,
eds. C.~Domb and M.~S.~Green (Academic Press, London 1974);
J.~Zinn-Justin, J. Physique 42 (1981) 783.

\bibitem{ItzyDrou}
C.~Itzykson and J.~M.~Drouffe, {\it in} Statistical field theory and lattice
gauge theory (Cambridge Univ. Press, Cambridge, 1989).

\bibitem{Adler}
J.~Adler, Y.~Meir, A.~Aharony and A.~B.~Harris, Phys. Rev.
B 41 (1990) 9183.

\bibitem{Leath}
P.~L.~Leath, Phys. Rev. B 14 (1976) 5064.

\bibitem{DeGennes}
P.~G.~DeGennes, La Recherche 7 (1976) 919.

\bibitem{GinsMoo}
G.~Moore and P.~Ginsparg, in Recent directions in particle theory,
eds. J.~Harvey and J.~Polchinski (World Scientific, Singapore, 1993)
pp. 537-588.

\bibitem{priv2}
A.~Coniglio, private communication.

\bibitem{co1}
A.~Coniglio, C.~R.~Nappi, F.~Peruggi and L.~Russo, \jpa 10 (1977)
205.

\end{thebibliography}
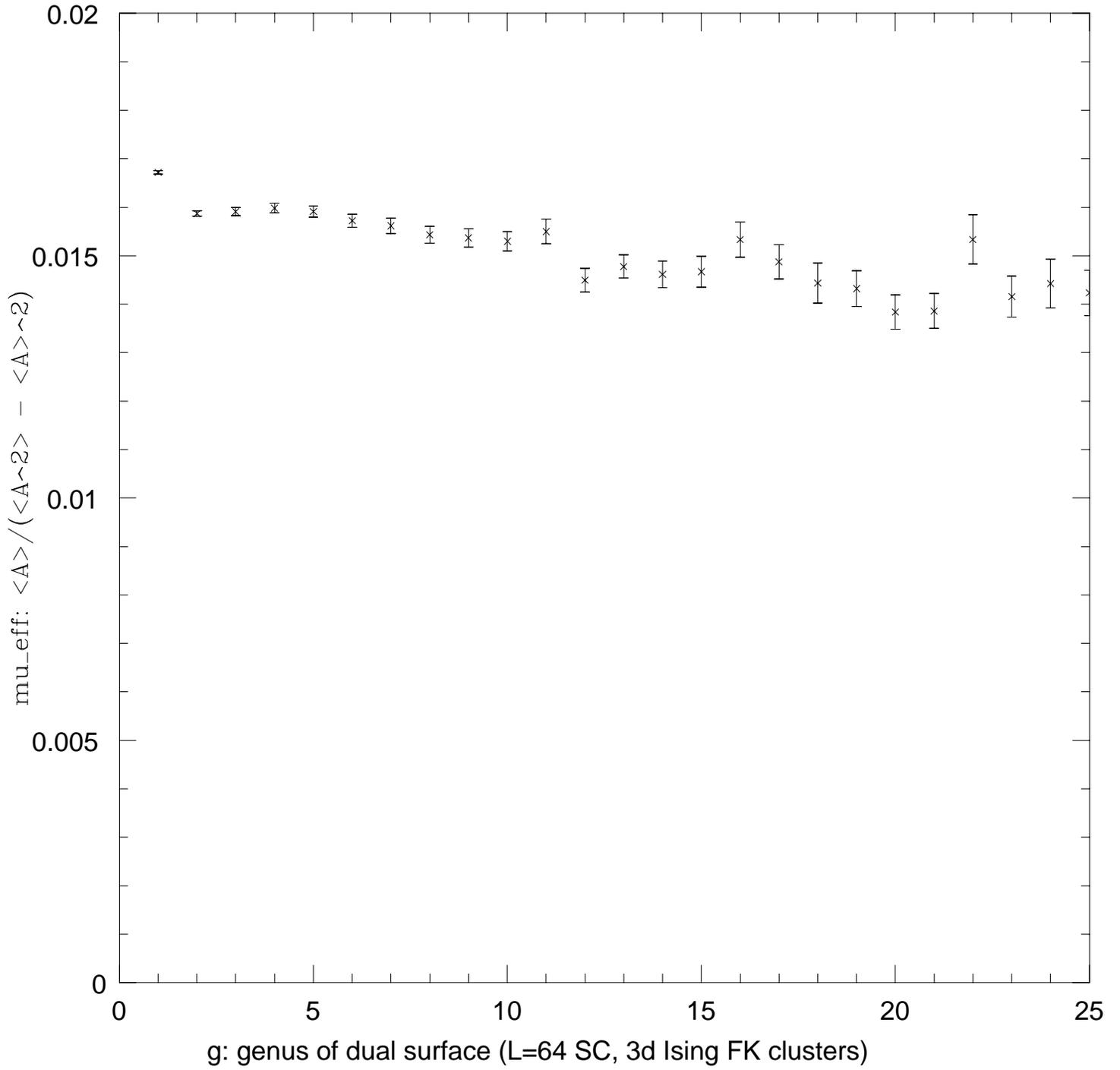
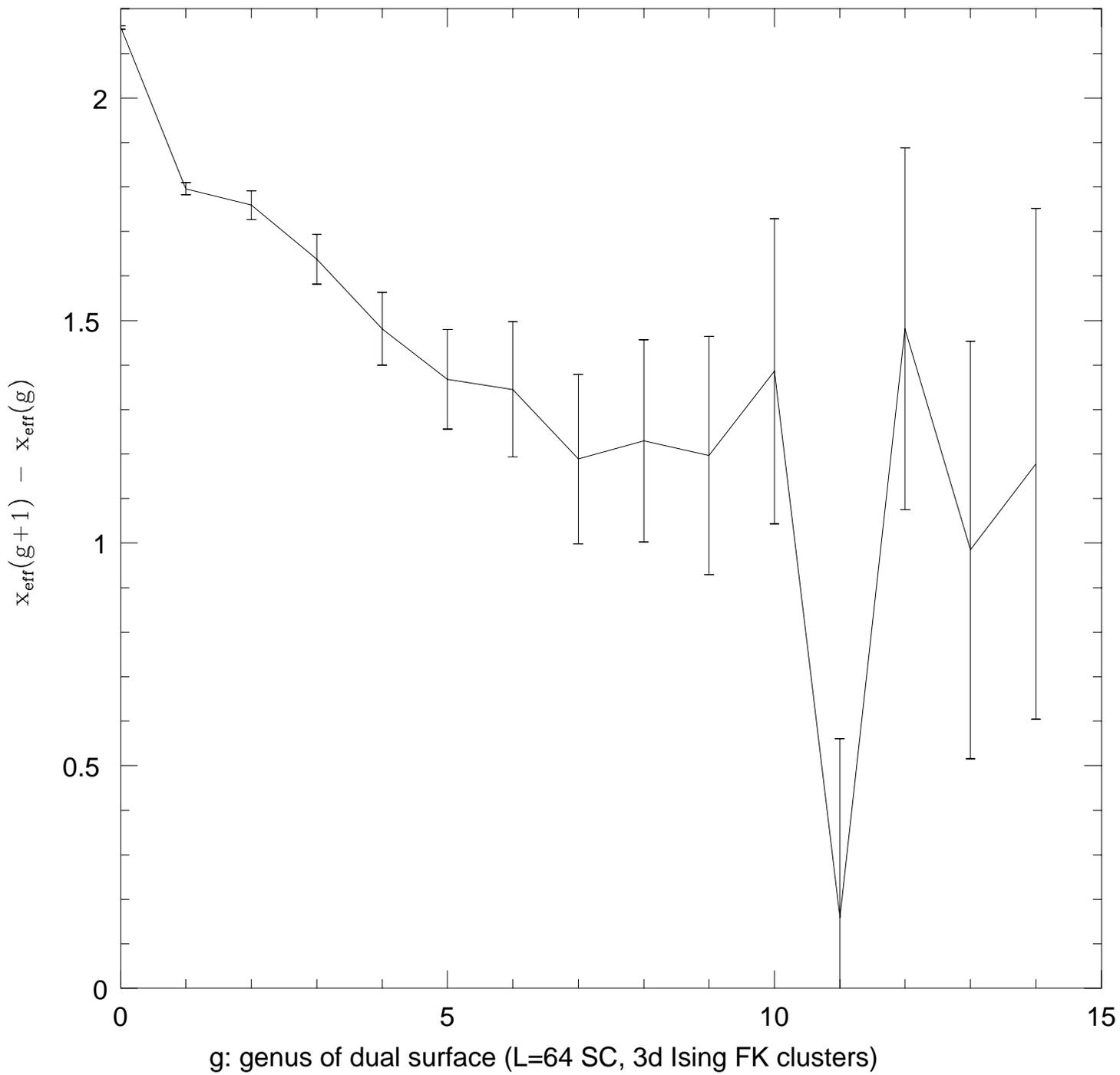
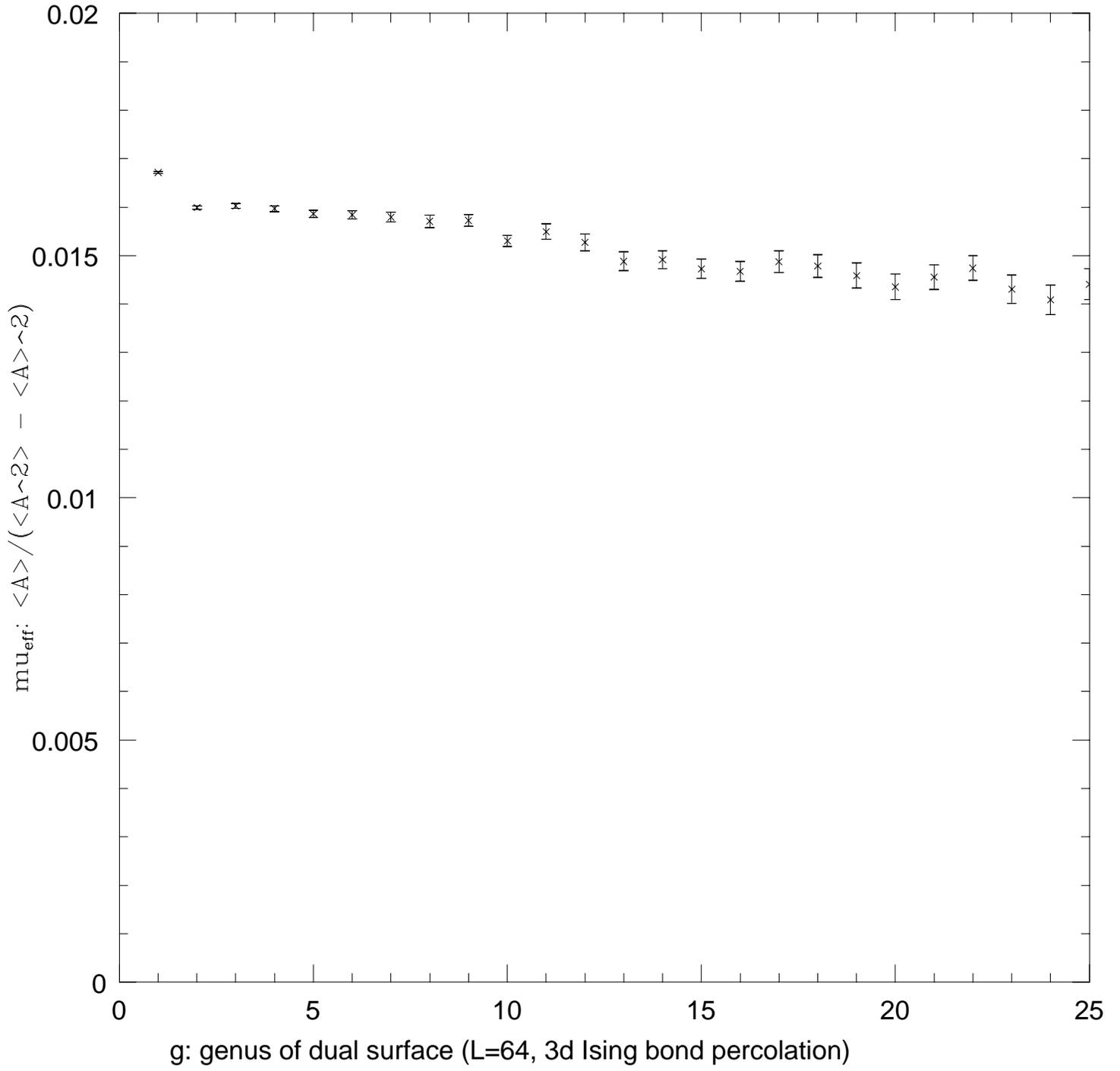
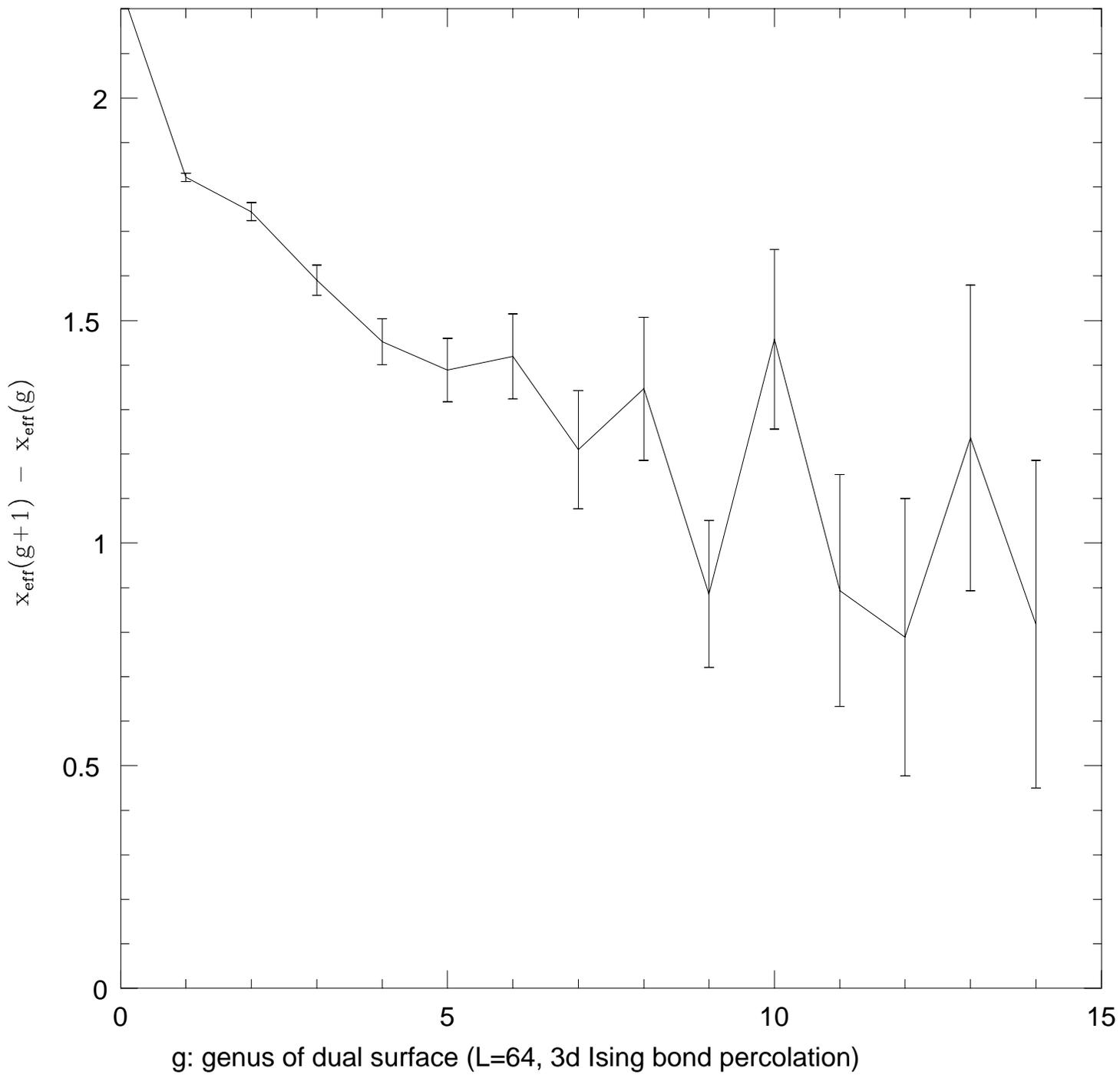
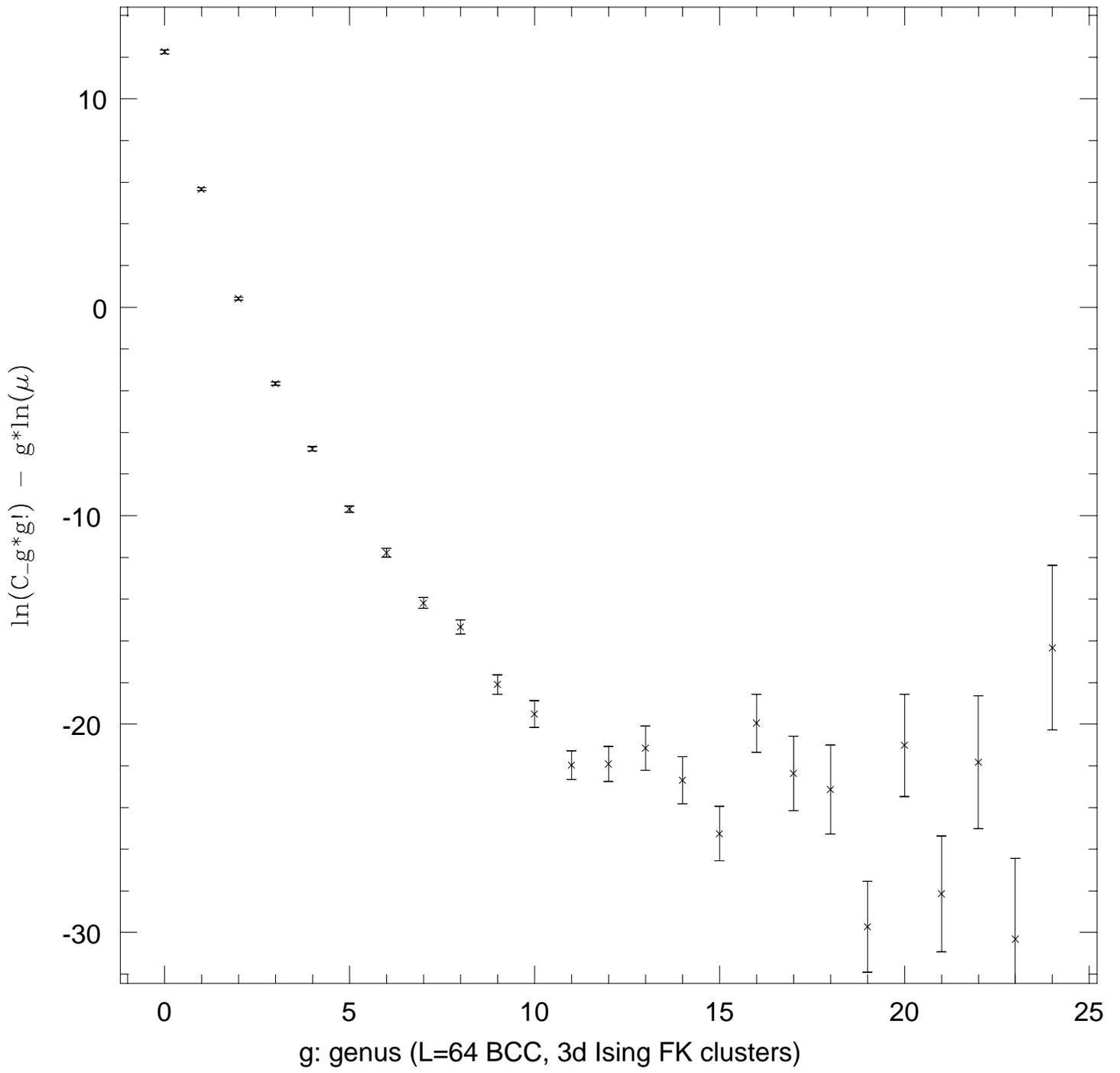
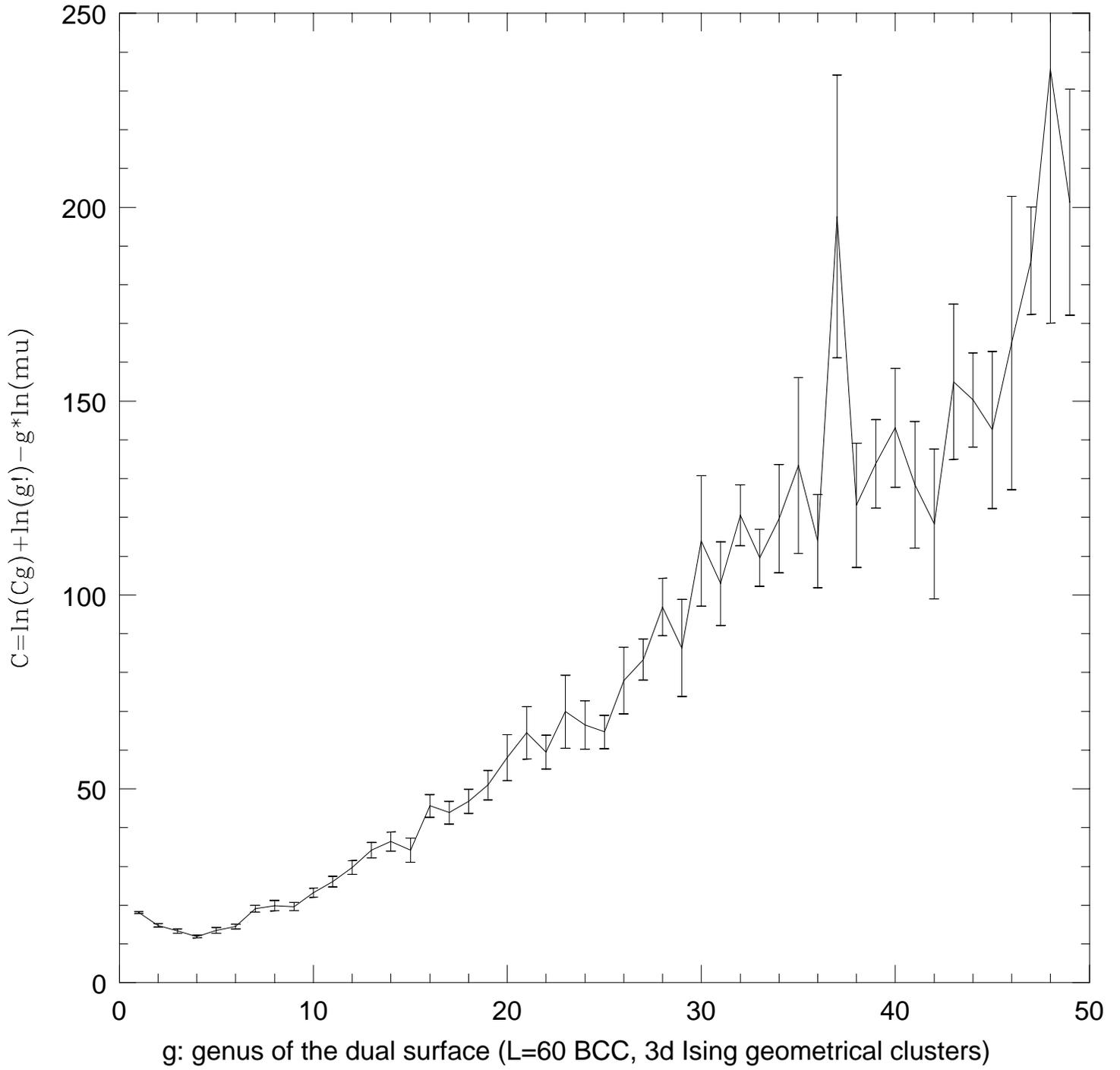
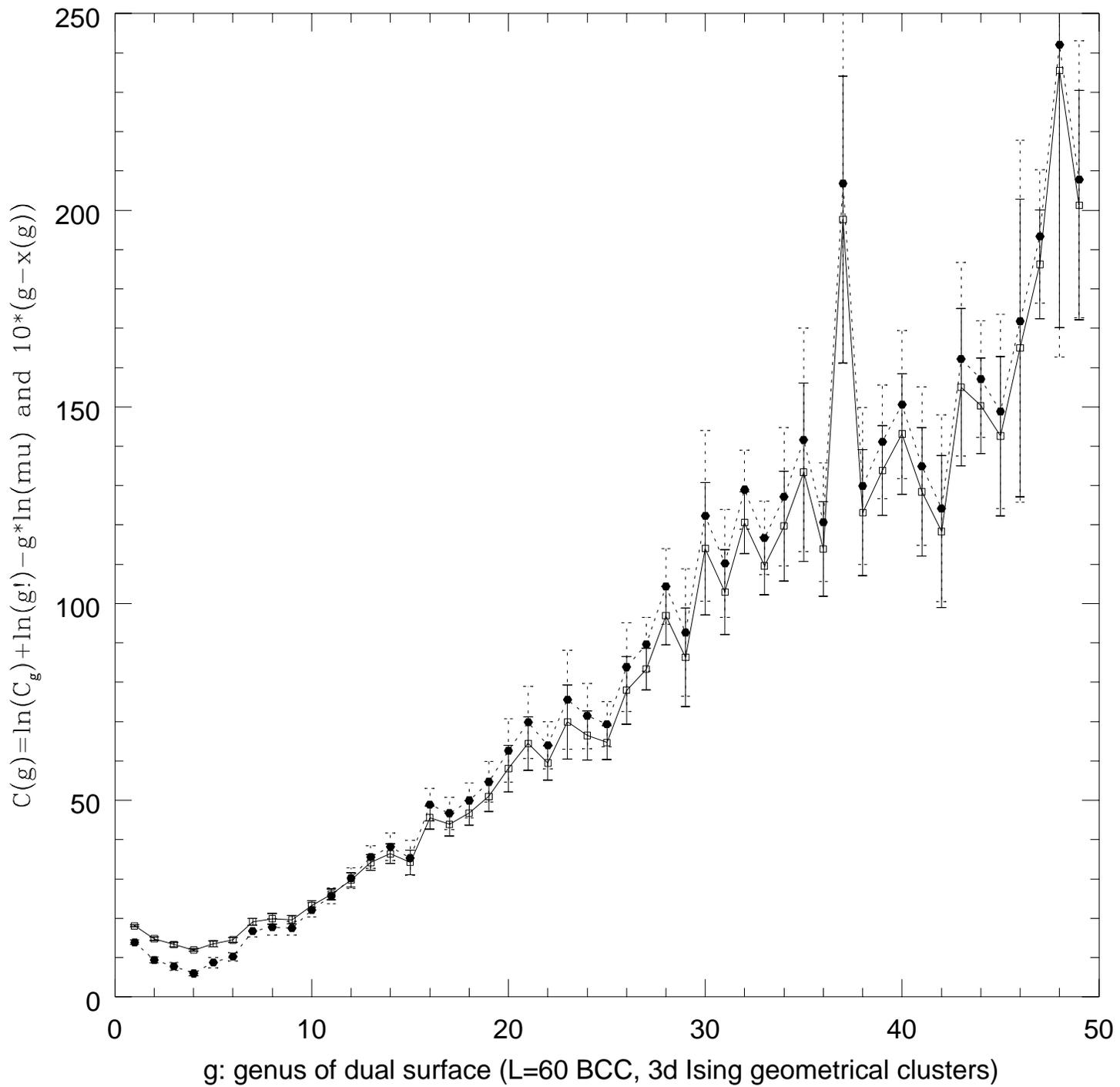
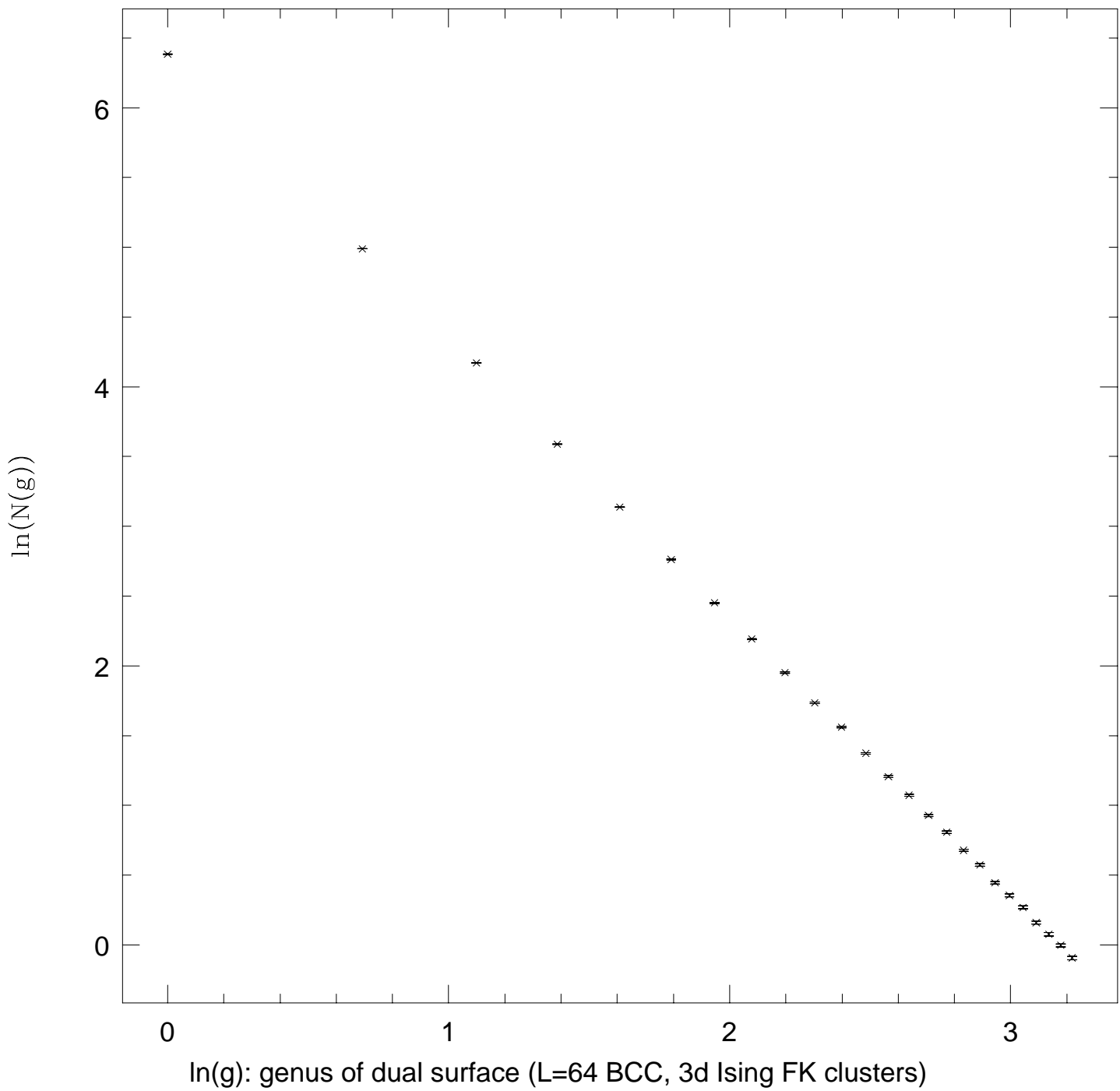
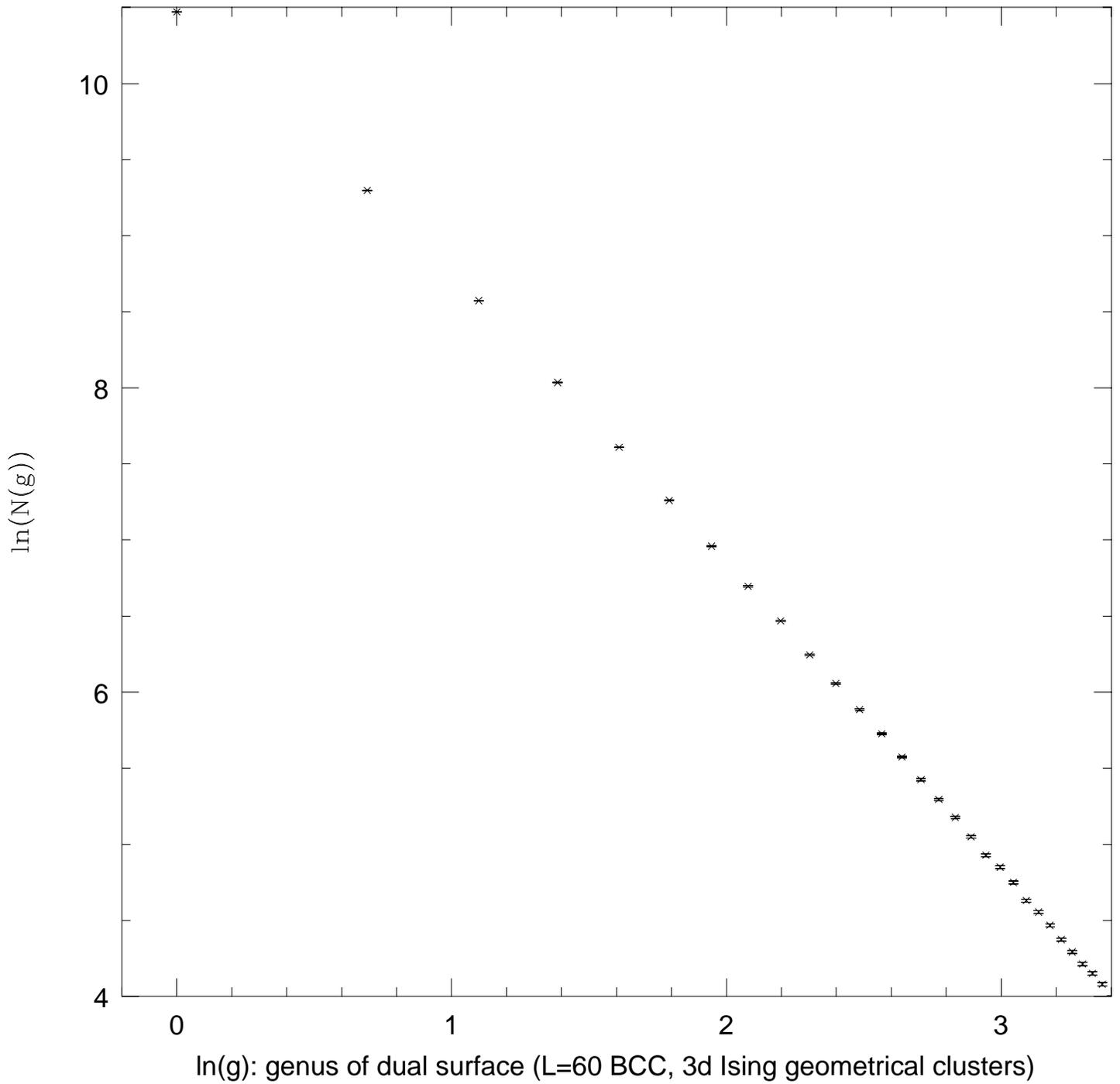
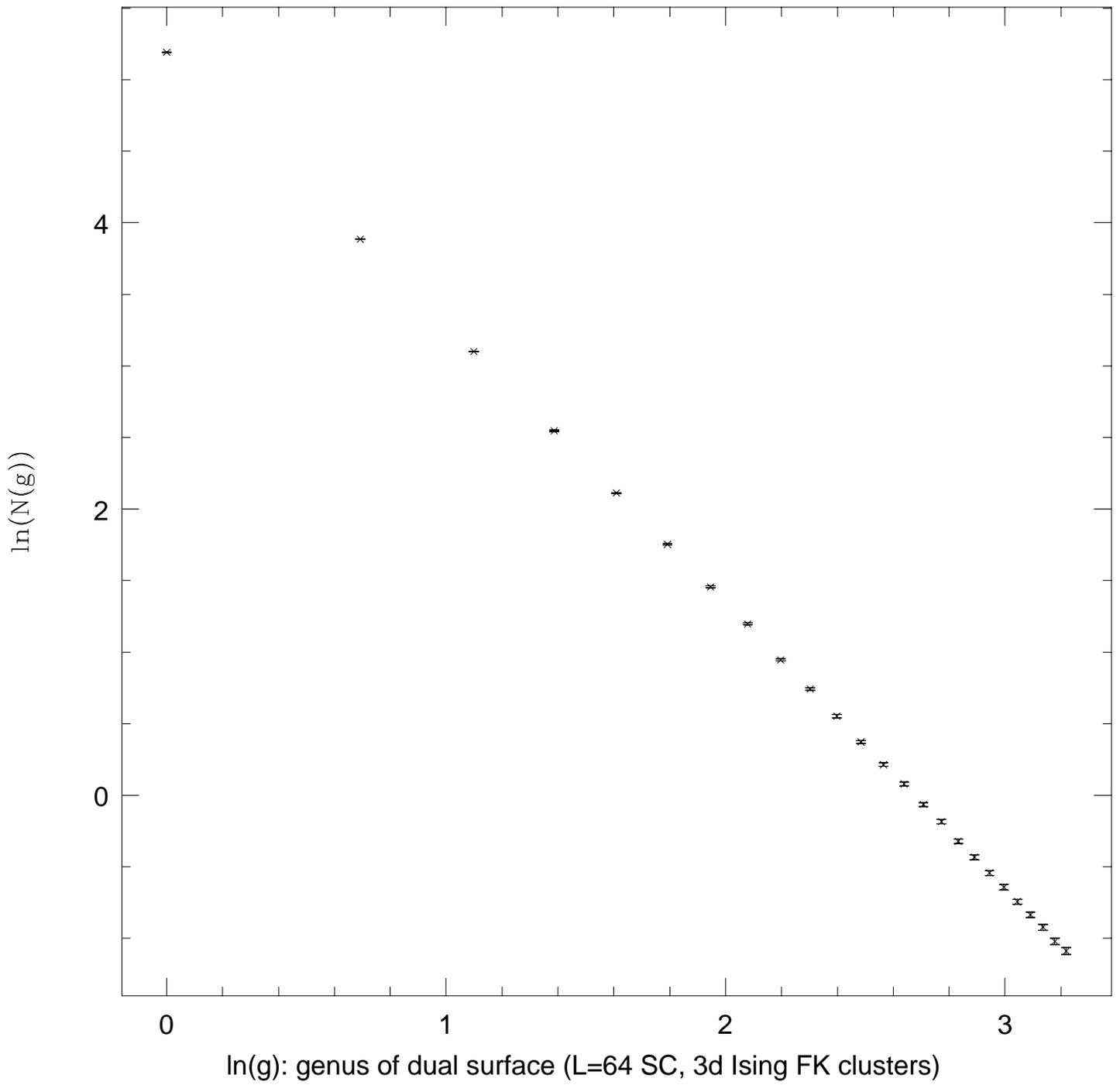
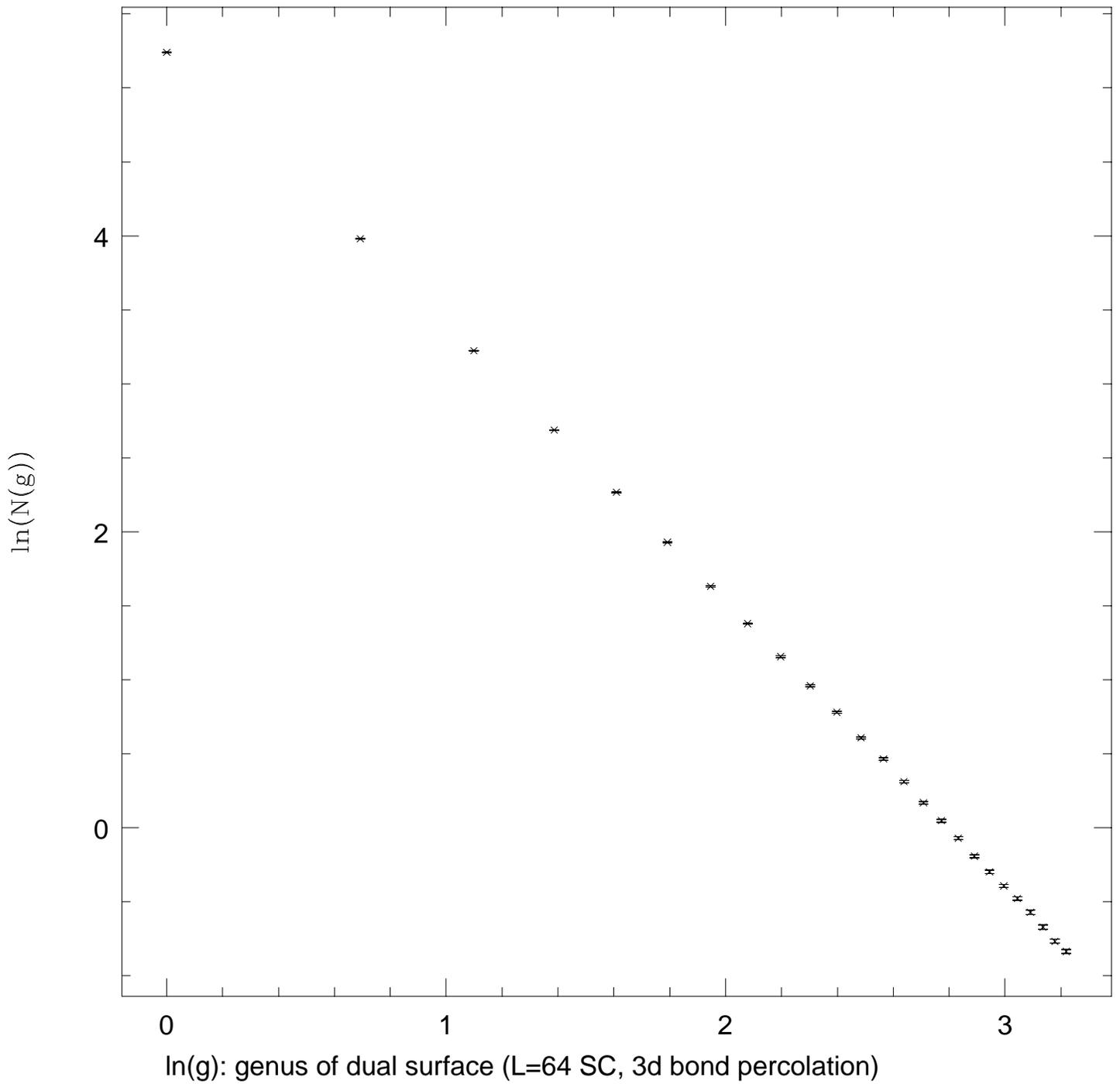
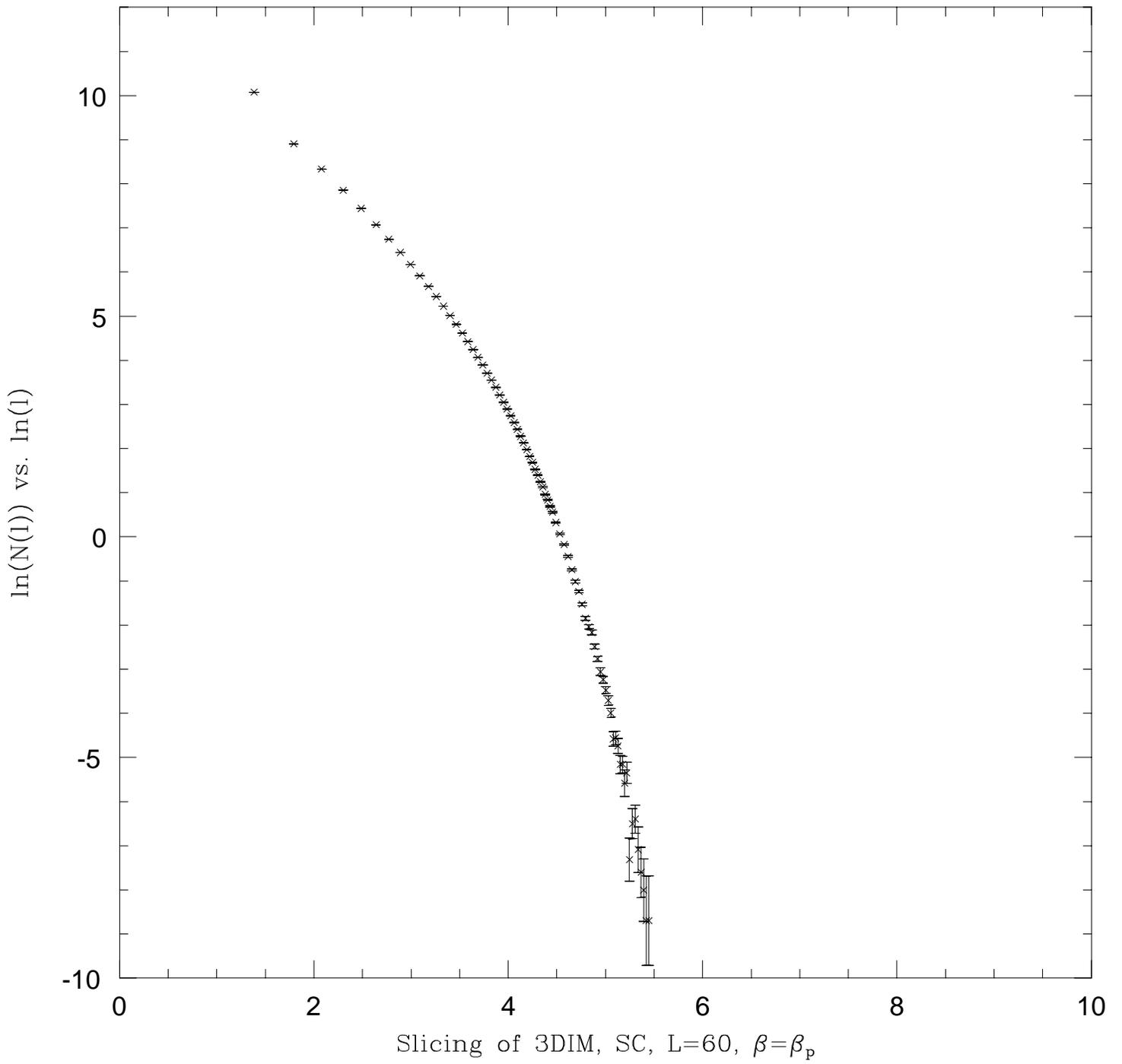
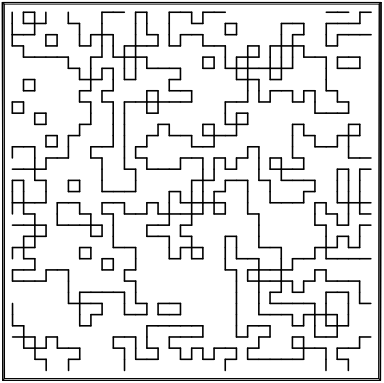
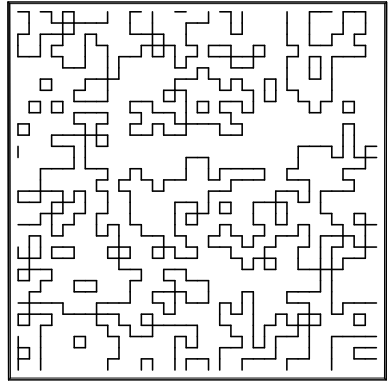
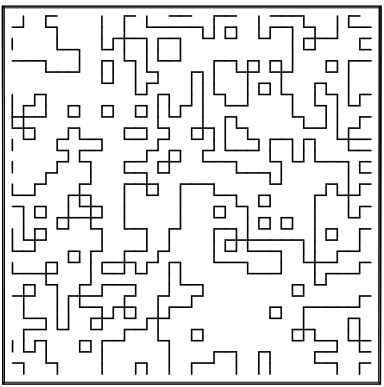
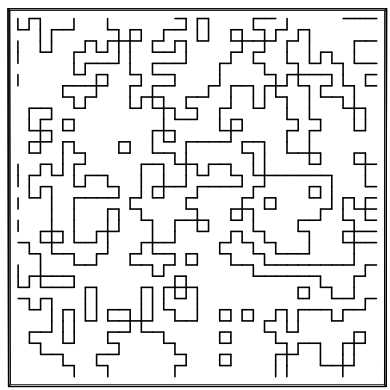
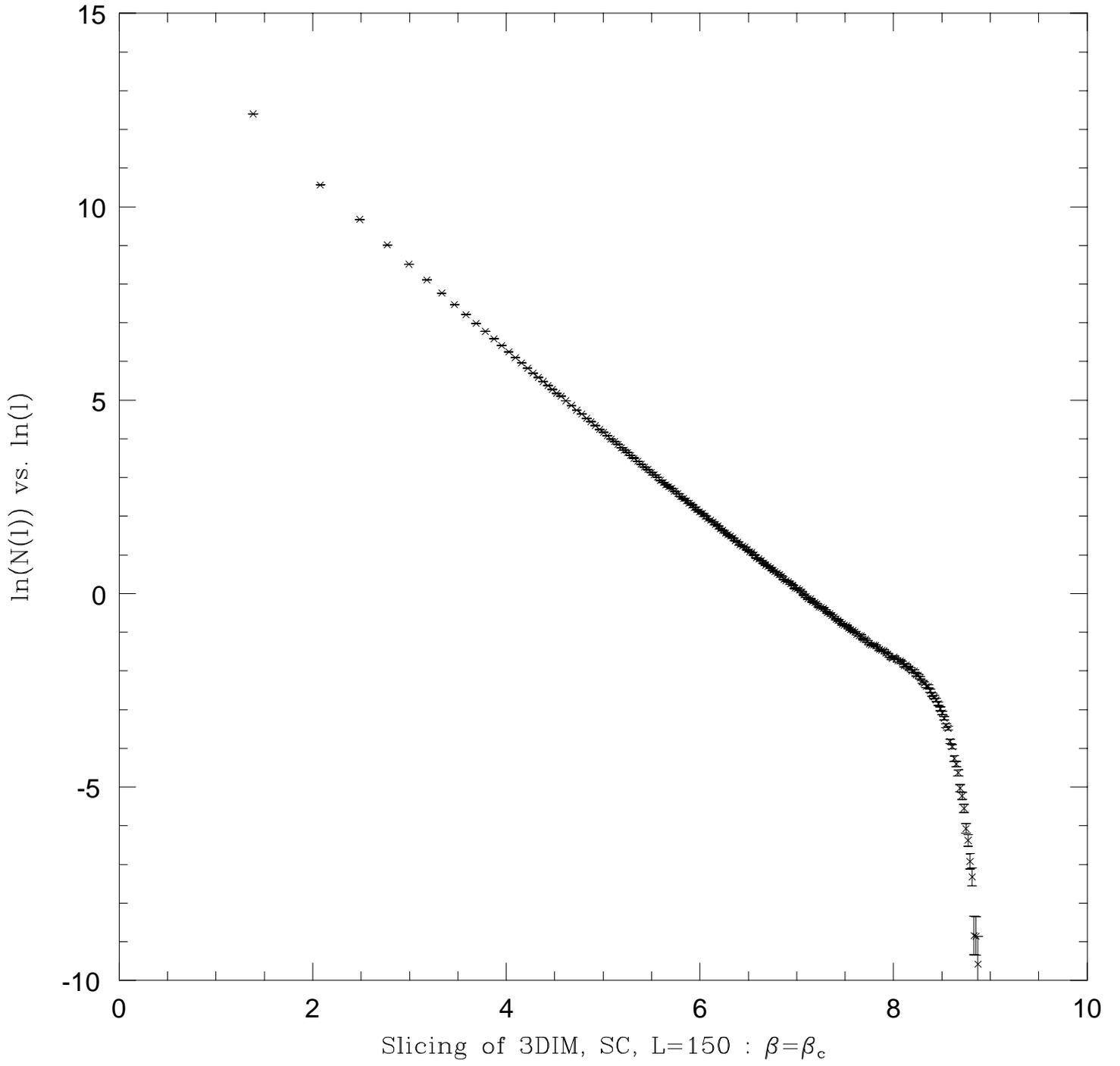
\end{document}